\documentclass{aa} 

\newcommand{\gtsim}{\protect\raisebox{-0.5ex}{$\:\stackrel{\textstyle >}
        {\sim}\:$}}
\newcommand{\ltsim}{\protect\raisebox{-0.5ex}{$\:\stackrel{\textstyle <}
        {\sim}\:$}}

\usepackage{graphicx}
\usepackage{aalongtable,lscape}   
\usepackage{txfonts}
\usepackage{rotating}
\usepackage{latexsym}
\usepackage{psfig}
\usepackage{epsfig}

\usepackage[authoryear]{natbib}
\bibpunct{(}{)}{;}{a}{}{,} 

\begin{document}

\title{Crossing the Gould Belt in the Orion vicinity\thanks{Based on ROSAT All-Sky Survey data, low-resolution spectroscopic 
observations performed at the European Southern Observatory (Chile; Program 05.E-0566) and at the 
{\it Observatorio Astron\'omico Nacional de San Pedro M\'artir} (M\'exico), and high-resolution spectroscopic 
observations carried out at the Calar Alto Astronomical Observatory (Spain).} 
\fnmsep\thanks{Figures \ref{fig:map} and \ref{fig:map_fe} are only available in electronic form at \texttt{http://www.aanda.org}.}
}
   
\author{K. Biazzo\inst{1} \and J. M. Alcal\'a\inst{1} \and E. Covino\inst{1} \and M. F. Sterzik\inst{2} \and P. Guillout\inst{3} 
 \and C. Chavarr\'{\i}a-K.\inst{4} \and A. Frasca\inst{5} \and R. Raddi\inst{6}}
\offprints{K. Biazzo}
\mail{katia.biazzo@oacn.inaf.it}

\institute{INAF - Osservatorio Astronomico di Capodimonte, via Moiariello, 16, 80131 Napoli, Italy
\and ESO - European Southern Observatory, Casilla 19001, Santiago 19, Chile
\and Observatoire Astronomique de Strasbourg, CNRS, UMR 7550, 11 rue de l'Universit\'e, 67000 Strasbourg, France
\and Instituto de Astronom\'{\i}a, Universidad Nacional Aut\'onoma de M\'exico, Ensenada, B. C., 22800, M\'exico
\and INAF - Osservatorio Astrofisico di Catania, via S. Sofia, 78, 95123 Catania, Italy
\and Centre for Astrophysics Research, STRI, University of Hertfordshire, College Lane, Hatfield AL10 9AB, United Kingdom}

\date{Received 02 November 2011 / Accepted 16 April 2012}

\abstract
{The recent star formation history in the solar vicinity is not yet well constrained, 
 and the real nature of the so-called Gould Belt is still unclear.}
{We present a study of the large-scale spatial distribution of 6482 ROSAT All-Sky Survey (RASS) 
X-ray sources in approximately 5000 deg$^2$  in the general direction of Orion. We examine the 
astrophysical properties of a sub-sample of $\sim 100$ optical counterparts, using optical 
spectroscopy. This sub-sample is then used to investigate the space density of the RASS young star 
candidates by comparing X-ray number counts with Galactic model predictions.
}  
{The young star candidates were selected from the RASS using X-ray criteria. We characterize 
the observed sub-sample in terms of spectral type, lithium content, 
radial and rotational velocities, as well as iron abundance. A population synthesis model is then
applied to analyze the stellar content of the RASS in the studied area.
} 
{We find that stars associated with the Orion star-forming region, as expected, do show a high 
lithium content. As in previous RASS studies, a population of late-type stars with lithium 
equivalent widths larger than Pleiades stars of the same spectral type (hence 
younger than $\sim 70-100$ Myr) is found widely spread over the studied area.
Two new young stellar aggregates, namely ``X-ray Clump 0534+22'' (age $\sim 2-10$ Myr) 
and ``X-ray Clump 0430$-$08'' (age $\sim 2-20$ Myr), are also identified.
}
{The spectroscopic follow-up and comparison with Galactic model predictions reveal that the X-ray 
selected stellar population in the general direction of Orion is characterized by three distinct 
components, namely the clustered, the young dispersed, and the widespread field populations. 
The clustered population is mainly associated with regions of recent or ongoing 
star formation and correlates spatially with molecular clouds. The dispersed young population 
follows a broad lane apparently coinciding spatially with the Gould Belt, while the widespread 
population consists primarily of active field stars older than 100 Myr. 
We expect the still ``bi-dimensional'' picture emerging from this study to grow in depth
as soon as the distance and the kinematics of the studied sources will become available from
the future {{\it Gaia}} mission.
}
   
	   
\titlerunning{Crossing the Gould Belt in the Orion vicinity}
\authorrunning{K. Biazzo et al.}
\maketitle

\section{Introduction}
If the global scenario of the star formation history (SFH) of the Milky Way is still not fully outlined (see \citealt{wyse2009} 
for a review), the recent SFH in the solar neighborhood is far from being constrained. In particular, the recent local star 
formation rate is poorly known because of the difficulty encountered in properly selecting young late-type stars in large sky 
areas from optical data alone (\citealt{guilloutetal2009}). This situation has improved thanks to wide-field X-ray observations,
such as the ROSAT\footnote{The R\"OentgenSATellit is an X-ray observatory that operated for nine years since the 
1st of June 1990, surveying the whole sky.} All-Sky Survey 
(RASS), which allowed for efficiently detection of young, coronally active stars in the solar vicinity (\citealt{guilloutetal1999}). 
Nearby star-forming regions (SFRs) have been searched for pre-main sequence (PMS) stars, and many new nearby moving groups and 
widely-spread young low-mass stars have been identified over the past decades based on the RASS data 
(\citealt{feigelsonmontmerle1999, zuckermansong2004, torresetal2008, guilloutetal2009}, and references therein). 

It has been suggested that at least some of these widely-spread young stars might be associated with the so-called Gould Belt 
(GB; \citealt{guilloutetal1998}), a disk-like structure made up of gas, young stars, and OB associations 
(see, e.g., \citealt{lindblad2000,eliasetal2009}). However, the existence of such structure and its possible origin remain 
somewhat controversial (\citealt{sanchezetal2007}, and references therein). Also, it is unclear whether its putative young 
stellar population is in excess with respect to predictions of Galactic models, which would indicate a recent episode of star 
formation. On the other hand, there is no evidence of molecular material within $\sim$100\,pc that can explain the origin of 
the distributed young stars as {\it in situ} star formation. Such young stars may thus represent a secondary star formation 
event in clouds that condensed from the ancient Lindblad ring supershell, or may have been formed in recent bursts of star 
formation from already disrupted and evaporated small clouds (see \citealt{bally2008}, and references therein).

An accurate census of the young population in different star-forming environments and the comparison between observations and 
predictions from Galactic models are thus required in order to constrain the low-mass star formation in the solar vicinity. 

In this work, we revisit the results by \cite{walteretal2000} on the large-scale spatial distribution of 6482 RASS X-ray sources
in a 5000~deg$^2$ field centered on the Orion SFR, and use optical, low and high resolution, spectroscopic observations to investigate 
the optical counterparts to 91 RASS sources (listed in Table~\ref{tab:parameters}) distributed on a sky ``strip" 
perpendicular to the Galactic Plane, in the general direction of Orion. The complete sky coverage of the RASS 
allows us to perform an unbiased analysis of the spatial distribution of X-ray active stars. Our main goal is to characterize 
spectroscopically the optical counterparts of the X-ray sources and compare their space density with predictions of Galactic models.
 
In Sect.~\ref{sec:x-ray_sel} we define the sky areas and describe the selection of young X-ray emitting star candidates; 
in Sect.~\ref{sec:obs} we present the spectroscopic observations; in Sect.~\ref{sec:follow-up} we characterize the
young optical counterparts in terms of spectral type, lithium detection, H$\alpha$ line, radial/rotational velocity 
and iron abundance; in Sect.~\ref{sec:discussion} we compare the spatial distribution of the young X-ray emitting star candidates 
with expectations from Galactic models; in Sect.~\ref{sec:conclusions} we discuss the results and draw our conclusions.

\section{Selection of study areas and X-Ray sources}
\label{sec:x-ray_sel}
Our analysis is based on the method introduced by \cite{sterziketal1995} for selecting young star candidates 
from the RASS sources using the X-ray hardness ratios and the ratio of X-ray to optical flux. Fig.~\ref{fig:map} is a 
revisited version of Fig.~3 by \cite{walteretal2000}, showing the color-coded space density of the RASS sources based 
on the $\alpha$ parameter, which is related to the probability of a source to be a young X-ray emitting star. 
Following the \cite{sterziketal1995} selection criteria, X-ray sources with $\alpha > 0.6$ are most likely young 
stars with a positive rate of 80\%. In the figure we recognize: 
$i)$~the surface density enhancements corresponding to 1483 young candidates, distributed over an area much larger 
than the molecular gas; 
$ii)$~space density enhancements not necessarily associated with previously known regions of active (or recent) 
star formation (e.g., at $\alpha=5^{\rm h}34^{\rm m}$, $\delta=+22{\degr}01'$ and at $\alpha=4^{\rm h}30^{\rm m}$, 
$\delta=-08{\degr}$); the local enhancement at $\alpha=5^{\rm h}07^{\rm m}$ and $\delta=-03{\degr}20'$ corresponds 
to the L1616 cometary cloud (\citealt{alcalaetal2004, gandolfietal2008}); 
$iii)$~a broad lane apparently connecting Orion and Taurus, which extends further southeastward; 
this wide and contiguous structure is not symmetric about the Galactic plane, but rather seems to follow 
the GB as drawn by \cite{guilloutetal1998}; 
$iv)$~the surface density of young star candidates drops down to a background value of about 
0.1 candidate star/deg$^2$ near $b_{II}=0{\degr}$, or below that value at higher Galactic latitudes.

\begin{figure*}{2}	
\begin{center}
 \begin{tabular}{c}
\includegraphics[width=14.5cm]{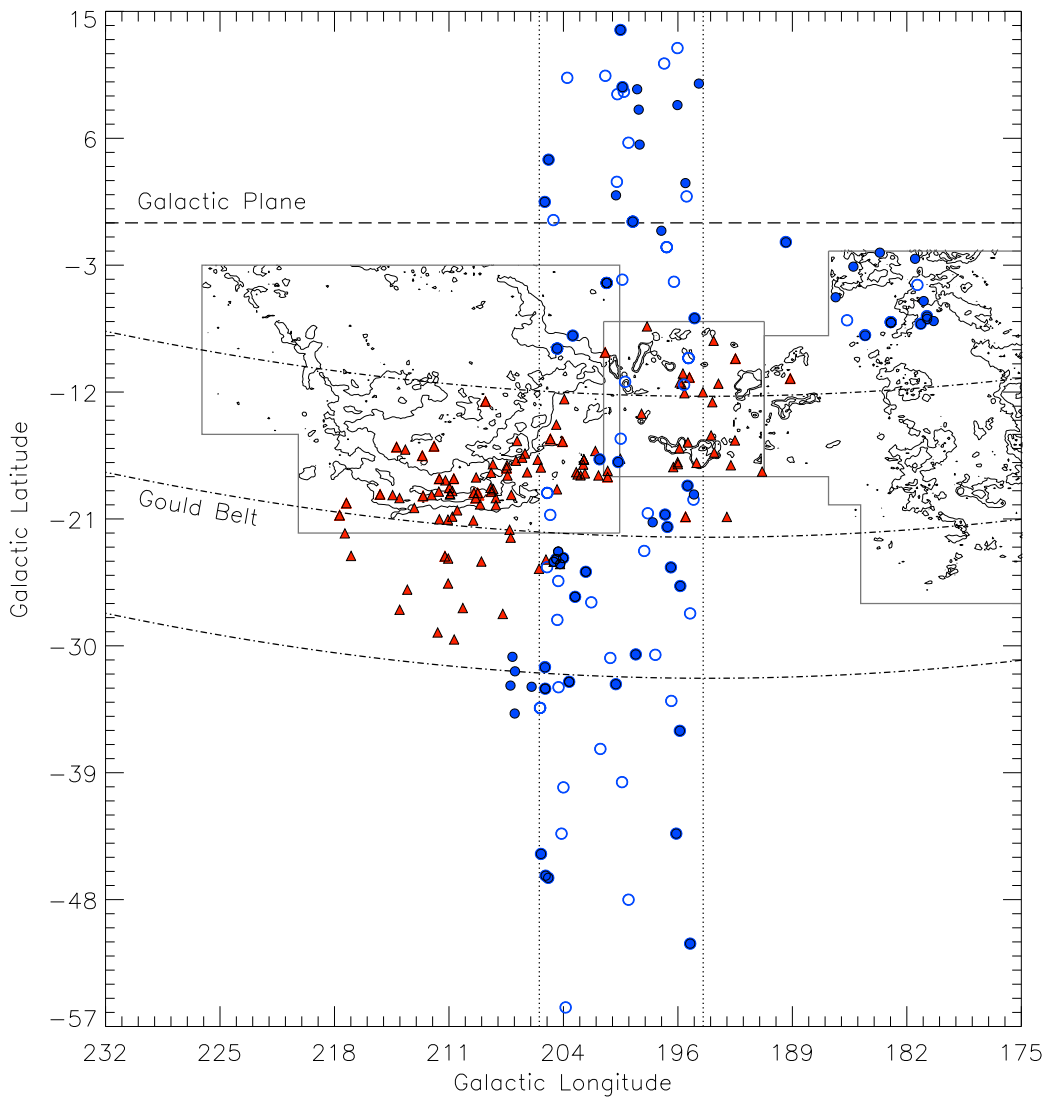}
 \end{tabular}
\vspace{-1cm}
\caption{Large-scale spatial distribution of our targets (circles) and those of Alcal\'a et al. (1996, 2000; triangles). Empty 
symbols refer to low-resolution spectra, while filled symbols represent objects observed at high resolution. The dotted lines 
mark the strip, the long-dashed line is the Galactic Plane, while the dash-dotted lines represent the Gould Belt disk, 
as outlined by \cite{guilloutetal1998}. The CO J=1$\rightarrow$0 emission contour maps by \cite{dameetal2001} of the Orion, 
Monoceros, and Taurus Molecular Clouds, and the $\lambda$~Orionis \ion{H}{ii} region are also overlaid.
} 
\label{fig:map_lowhigh}
 \end{center}
\end{figure*}

In order to study the global properties of the different populations close to the Orion SFR, we selected a 
$10{\degr}\times75{\degr}$ strip (see Fig.~\ref{fig:map}) perpendicular to the Galactic Plane (hence presumably crossing the 
GB), and in the Orion vicinity. Inside this $\sim 750$ deg$^2$ strip ($l_{II}^{\rm min}=195{\degr}$, 
$l_{II}^{\rm max}=205{\degr}$, $b_{II}^{\rm min}=-60{\degr}$, $b_{II}^{\rm max}=15{\degr}$), 806 RASS X-ray sources were 
detected with a high confidence level by the Standard Analysis Software System (SASS; \citealt{voges1992}). According to the 
\cite{sterziketal1995} selection criteria, 198 of these sources are young star candidates. Additionally, we selected a 
region of $10{\degr}\times10{\degr}$ (see Fig.~\ref{fig:map}) centered at $\alpha=5^{\rm h}34^{\rm m}$ and 
$\delta=+22{\degr}01'$, where a density enhancement of young star candidates is present. We call this previously 
unrecognized enhancement ``X-ray Clump 0534+22'' (hereafter, X-Clump 0534+22). Analogously, we also identify as 
``X-ray Clump 0430$-$08'' (hereafter, X-Clump 0430$-$08) the unknown density enhancement at $\alpha=4^{\rm h}30^{\rm m}$ 
and $\delta=-08{\degr}$. Although the X-Clump~0534+22 almost coincides in direction with the Crab nebula, it 
is physically unrelated to the supernova remnant.

\section{Spectroscopic observations and data reduction}
\label{sec:obs}
We conducted a spectroscopic follow-up of 91 young stellar candidates inside the strip and clump regions.
Table~\ref{tab:logs} gives a summary of the spectroscopic observations, while the full list of the observed 
stars is reported in Table~\ref{tab:parameters}). Throughout the paper, we term as `young star' an optical 
counterpart that shows characteristics typical of weak-lined T-Tauri stars, i.e. weak H$\alpha$ emission 
($EW_{\rm H\alpha} \ltsim 10$ \AA) and strong lithium absorption. 

\subsection{Low-resolution spectroscopy}
\label{sec:low-res}
Low-resolution spectroscopic observations were carried out during 25-30 November 1995 and 16-21 December 1996 using the Boller 
\& Chivens (B\&Ch) Cassegrain spectrographs attached to the 1.5m telescope of the European Southern Observatory (ESO, Chile) 
and to the 2.1m of the {\it Observatorio Astron\'omico Nacional de San Pedro M\'artir} (OAN-SPM, M\'exico), respectively. 
Table~\ref{tab:logs} gives information on the instrumental setups and number of observed objects. The spectral resolution 
was verified by measuring the full width at half maximum of several lines in calibration spectra. The spectra were reduced 
following the standard procedure of MIDAS\footnote{The MIDAS (Munich Image Data Analysis System) system provides general tools 
for image processing and data reduction. It is developed and maintained by the ESO.} software packages using the same procedure 
described in \citet{alcalaetal1996}.

\setlength{\tabcolsep}{4pt}
\begin{table}
\caption{Summary of the spectroscopic observations.} 
\label{tab:logs}
\begin{center}
\begin{tabular}{llcccc}
\hline
\hline
Telescope 	& Instrument &   Range & Resolution             & \#\\ 
 		&      	      & (\AA)& ($\lambda/\Delta\lambda$) & stars\\
\hline
1.5m@ESO     & B\&Ch &  3400--6800 & 1\,600  &	66   \\
2.1m@OAN-SPM & B\&Ch &  3600--9900 & 1\,500  &	6 \\
2.2m@CAHA    & FOCES &  4200--7000 & 30\,000 &  61  \\
\hline
\end{tabular}
\end{center}
\footnotesize{Note: 30 stars were observed only at low resolution, and 20 only at high 
resolution. One star was observed at low resolution with both the B\&Ch spectrographs.} 
\end{table}

About sixty of the 198 strip sources plus fourteen stars in the X-Clump 0534+22 direction were investigated 
spectroscopically at low resolution (see Table~\ref{tab:parameters} and Fig.~\ref{fig:map_lowhigh}). The observational strategy 
consisted in covering the whole range of right ascension and declination each night in order to avoid any bias in the resulting 
spatial distribution of the young star candidates. The spatially unbiased sample so far observed and characterized spectroscopically 
by us is therefore incomplete, representing $\sim$40\% of the total sample of potential young X-ray emitting candidates. Yet, 
it can be used to study the strength of the lithium absorption line within the RASS young stellar sample as a function of 
Galactic latitude, and to trace the young stellar population in the general direction of Orion.  

\subsection{High-resolution spectroscopy}
\label{sec:high-res}
High-resolution spectroscopic observations were conducted in several runs in the period between October 1996 and December 
1998, using the Fiber Optics Cassegrain \'Echelle Spectrograph (FOCES) attached to the 2.2m telescope at the Calar Alto 
Observatory (CAHA, Spain). Some seventy spectral orders are included in these spectra covering the range from 4200 to 7000~\AA, 
with a nominal mean resolving power of $\lambda/\Delta\lambda\approx$30\,000 (see Table~\ref{tab:logs}). The reduction was performed 
using IDL\footnote{IDL (Interactive Data Language) is a registered trademark of ITT Visual Information Solutions.} routines 
specifically developed for this instrument (\citealt{pfeifferetal1998}). Details on the data reduction are given in 
\citet{alcalaetal2000}.

We also retrieved FEROS\footnote{This is the Fiber-fed Extended Range Optical Spectrograph operating in La Silla (ESO, Chile) 
for the 1.5m telescope.} spectra for two stars of our sample (namely, 2MASS~J03494386$-$1353108 and 2MASS~J04354055$-$1017293) from 
the ESO Science Archive\footnote{http://archive.eso.org/cms/}. The FEROS spectra extend between 3600 \AA~and 9200 \AA~with 
a resolving power $R$=48\,000 (\citealt{kauferetal1999}). The data were reduced using a modified version of the FEROS-DRS 
pipeline (running under the ESO-MIDAS context FEROS) which yields a wavelength-calibrated, merged, normalized spectrum, 
following the steps specified in \citet{desideraetal2011}.

In summary, high-resolution spectroscopy exists for 61 stars, 33 inside inside the strip, 13 in the X-Clump 0534+22, 
8 associated with the high space-density X-Clump 0430$-$08 (see in Fig.~\ref{fig:map} the spatial distribution at 
$\alpha=4^{\rm h}30^{\rm m}$ and $\delta=-08{\degr}$), and 7 inside the strip associated to the L1616 clump. 
The high-resolution sample thus represents $\sim$70\% of the known lithium stars in the strip, and can 
be used to verify the reliability of the lithium strength obtained from the low-resolution spectra.

In order to increase the statistics, we also include in our analysis the on-strip lithium-rich stars identified by 
\cite{alcalaetal1996} and \cite{alcalaetal2000}, which were observed with the same instruments, set-ups, and observational 
strategy.

\section{Characterization of the selected young X-ray counterparts}
\label{sec:follow-up}
\subsection{Near-IR color-color diagram}
\begin{figure}  
\begin{center}
 \begin{tabular}{c}
\hspace{-1cm}
\includegraphics[width=10cm]{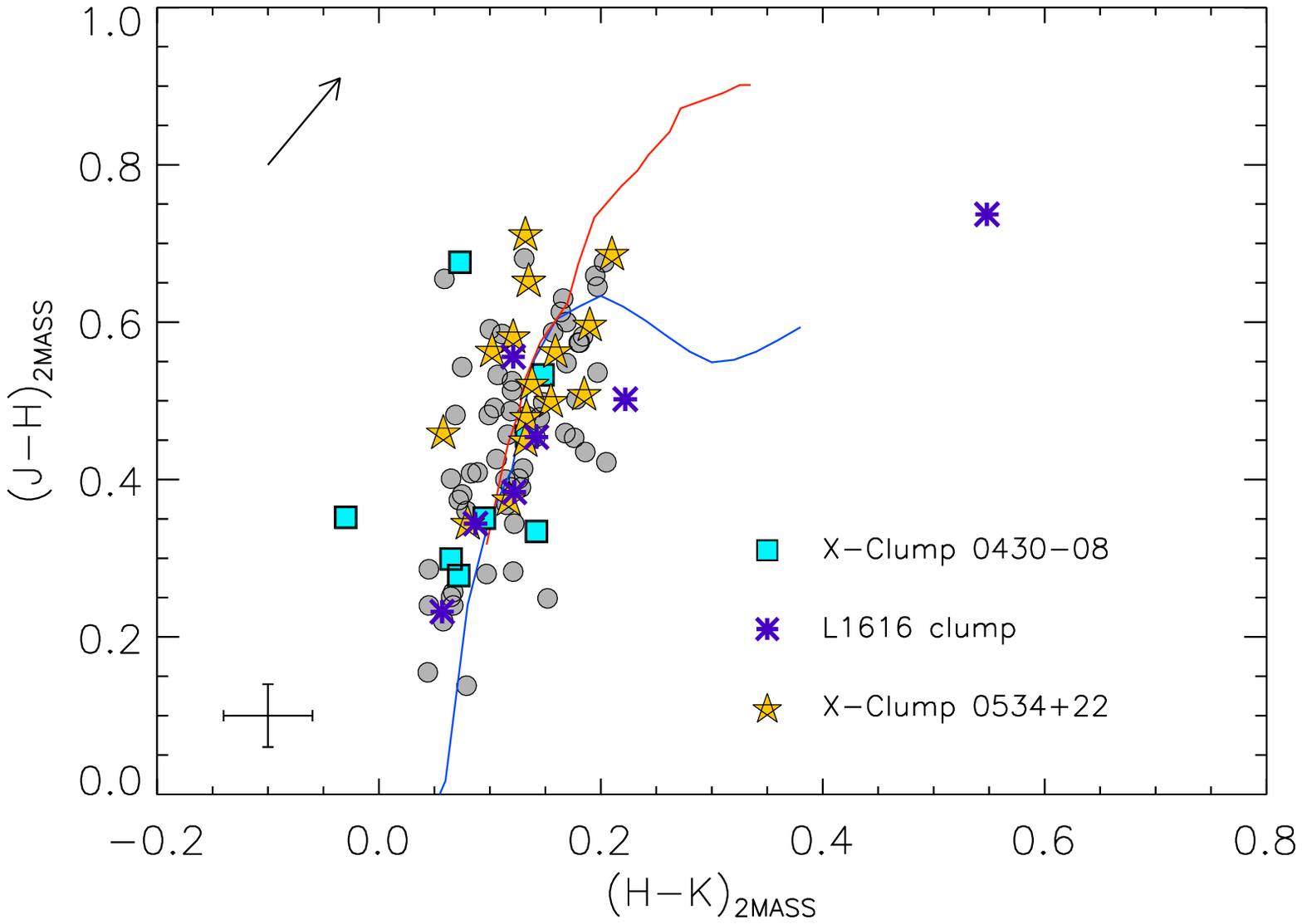}
 \end{tabular}
\caption{{\it 2MASS} $(J-H)$ versus $(H-K)$ diagram for the observed sources. Squares, asterisks, and stars represent objects in 
the three clumps, as indicated in the legend. The circles refer to the ``non-clump'' stars. The solid curve shows the relation 
between these indexes for main sequence stars (lower branch; \citealt{bessellbrett1988}) and giants (upper branch; 
\citealt{kenyonhartmann1995}), where the {\it 2MASS} color trasformations were used. The $A_{\rm V}=1$ mag reddening vector 
is shown with an arrow. The typical {\it 2MASS} photometric errors are overplotted on the lower-left corner of the panel.
} 
\label{fig:ccd_2mass}
 \end{center}
\end{figure}

We identified all our targets in the {\it Two Micron Sky Survey} ({\it 2MASS}) catalogue. In Table~\ref{tab:parameters} we list both 
the RASS Bright Source Catalogue (\citealt{vogeset1999}) and {\it 2MASS} designations, as well as an alternative name. The stellar 
coordinates are those from the {\it 2MASS} catalogue. We then examined the properties of the sources in the $JHK$ bands using their 
{\it 2MASS} magnitudes (\citealt{cutrietal2003}) looking for eventual color excess. The  $(J-H)$ versus $(H-K)$ diagram 
(Fig.~\ref{fig:ccd_2mass}) shows that our sample mostly consists of stars without near-IR excess, 
with the only exception of  2MASS~J05122053$-$0255523=V531~Ori, which was classified as a classical T Tauri star by 
\cite{gandolfietal2008}. All stars follow the MS branch with a spread mostly ascribable to photometric uncertainties. Only 
2MASS~J04405981$-$0840023 departs from the sequence, maybe due to its double-lined binary nature (\citealt{covinoetal2001}).

\subsection{Spectral types, lithium detection, and H$\alpha$ equivalent width}
\label{sec:sptype_lithium_halpha}
\begin{figure}	
\begin{center}
 \begin{tabular}{c}
\hspace{-1cm}
\includegraphics[width=10cm]{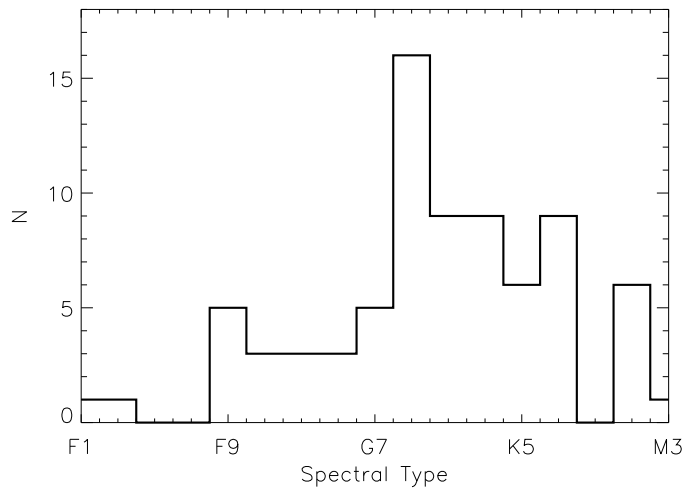}
 \end{tabular}
\caption{Spectral type distribution of the young stellar candidates.} 
\label{fig:sptypes}
 \end{center}
\end{figure}

Spectral types were determined from the low-resolution spectra by comparison with a grid of bright spectral standard stars (from F0 to M5) 
observed with the same dispersion and instrumental set-up in each observing run. The methods described in \cite{alcalaetal1995} were 
used for the classification, leading to an accuracy of about $\pm 1$ sub-class in most cases. The spectral types are reported in 
Table~\ref{tab:parameters}, while their distribution is plotted in Fig.~\ref{fig:sptypes}. The sample is composed of late-type stars 
with a distribution peaked around G9--K1.

In our low-resolution spectroscopic follow-up, the Orion stars fall basically in four categories (see Table~\ref{tab:parameters}): 
$i)$ stars with weak H$\alpha$ emission ($-3$~\AA$\ltsim EW_{\rm H\alpha} \ltsim 0$~\AA) and Li absorption (21 stars); 
$ii)$ stars with H$\alpha$ filled-in or in absorption and Li  absorption (40 stars); $iii)$ stars with H$\alpha$ in emission 
but no Li absorption (8 stars); and $iv)$ stars with H$\alpha$ in absorption but no Li detection (4 stars). In total, 61 stars with clear 
lithium detections were found throughout the strip, as well as in the clumps. Practically all lithium stars have spectral types 
ranging from late F to K7/M0 peaking around G9 (see Table~\ref{tab:parameters}). The effective temperature versus spectral type 
relation for dwarfs by \cite{kenyonhartmann1995} was used to estimate the $\log T_{\rm eff}$ values listed in 
Table~\ref{tab:parameters}. 

At this point, it is important to stress that: $i)$ at the Orion distance, our sample is limited to masses $> 0.8M_{\odot}$ 
(\citealt{alcalaetal1998}) because of the RASS flux limit; $ii)$ the Li equivalent width 
($EW_{\rm Li}$) may be overestimated in low-resolution spectra because of blending mainly with the nearby \ion{Fe}{i} line 
at $\lambda=6707.4$ \AA~(see Sect.~\ref{sec:lithium_high-res}). The latter issue can be overcome by using high-resolution 
spectroscopy (see Sections \ref{sec:high-res} and \ref{sec:lithium_high-res}).

\subsection{Lithium strength: low- versus high-resolution measurements}
\label{sec:lithium_high-res}
For 43 stars we obtained both low- and high-resolution spectra. Based on these data, we estimated that the mean 
lithium equivalent width measured on low-resolution spectra ($EW_{\rm Li}^{\rm lr}$) is overestimated by 
$\sim$20 m\AA, with a standard deviation of $\pm60$ m\AA~on the average difference between the low and high resolution 
measurements. Interestingly, for several stars, the $EW_{\rm Li}^{\rm lr}$ matches well with the lithium 
equivalent width obtained from high-resolution spectra ($EW_{\rm Li}^{\rm hr}$) and, in some cases, $EW_{\rm Li}^{\rm lr}$ 
values are underestimated (see Table~\ref{tab:parameters}). The good match between the $EW_{\rm Li}^{\rm lr}$ and 
$EW_{\rm Li}^{\rm hr}$ values is due to the fact that the lithium strength of these stars is indeed high and also 
to the experience we gathered on measuring $EW_{\rm Li}$ in low-resolution spectra.

\begin{figure*}
\begin{center}
 \begin{tabular}{c}
\includegraphics[width=9.2cm]{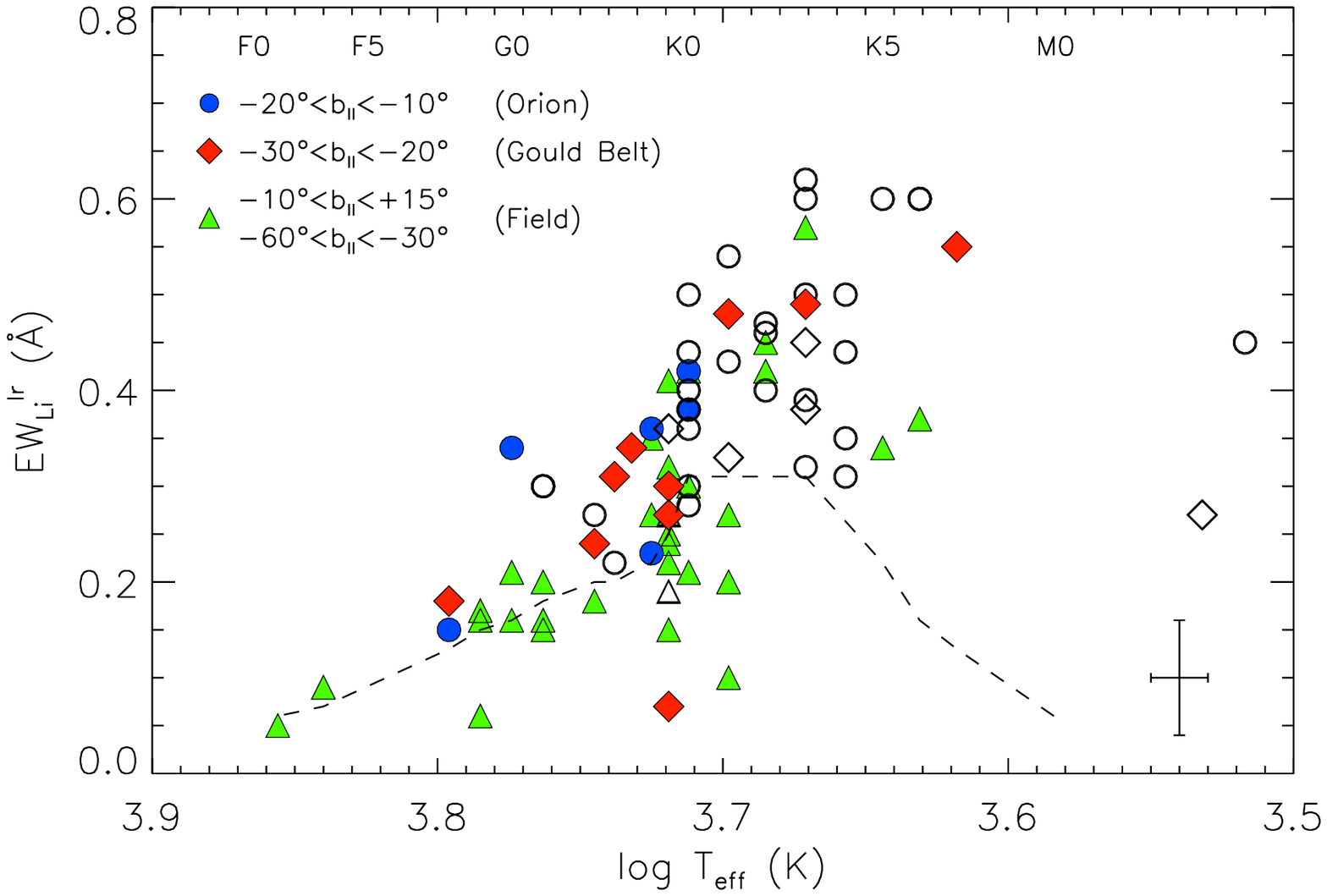}
\includegraphics[width=9.2cm]{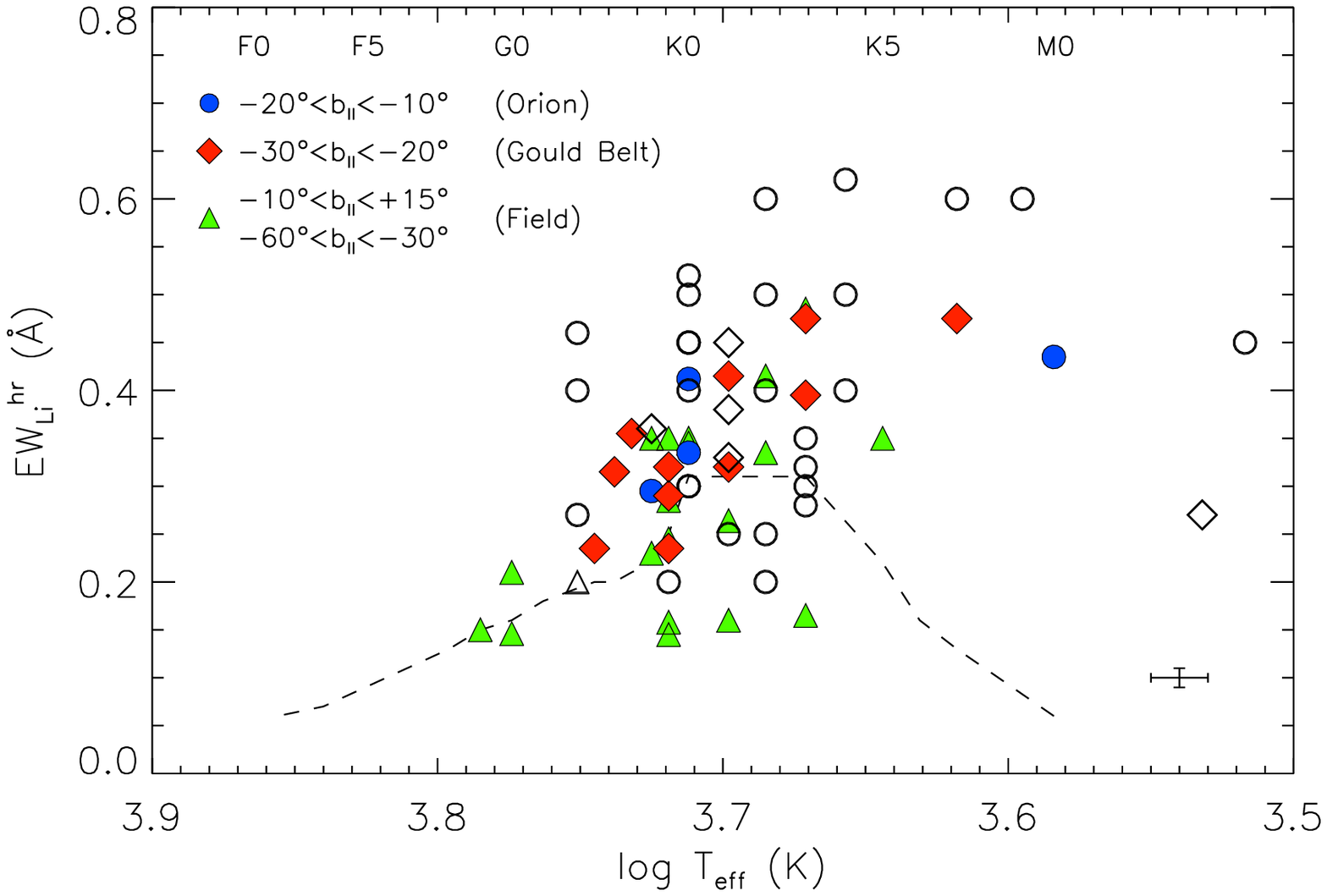}
 \end{tabular}
\caption{Lithium ($\lambda$6707.8 \AA) equivalent width versus $\log T_{\rm eff}$ for the stars in the strip 
($195\degr<l_{II}<205\degr$), observed at low ({\it left panel}) and high ({\it right panel}) resolution. Three different 
bins of Galactic latitude were considered (see also Fig.~\ref{fig:map_lowhigh}). The upper envelope for the Pleiades sample, as adapted 
from \cite{soderblometal1993}, is overplotted as a dashed line. Mean errors in $EW_{\rm Li}^{\rm lr}$ ($\sim$60 m\AA), 
$EW_{\rm Li}^{\rm hr}$ ($\sim$10 m\AA), and $\log T_{\rm eff}$ ($\sim$100 K, as obtained from the spectral synthesis) are 
overplotted on  the lower-right corner of each panel. Open symbols represent the \cite{alcalaetal1996} and \cite{alcalaetal2000} 
results from low ({\it left panel}) and high ({\it right panel}) resolution spectra, respectively.} 
\label{fig:teffli}
 \end{center}
\end{figure*}

In Fig.~\ref{fig:teffli} (left panel), the lithium equivalent widths of the 47 stars in our sample and the 
38 from \cite{alcalaetal1996}, observed at low-resolution and falling inside the strip, are plotted versus the 
logarithm of the effective temperature, for the following three bins of Galactic latitude: $i)$ $-20{\degr}<b_{II}<-10{\degr}$, 
coinciding with the Orion Complex, $ii)$ $-30{\degr}<b_{II}<-20{\degr}$, corresponding approximately to the 
position of the Gould Belt at those Galactic longitudes, and $iii)$ in the two ranges $-10{\degr}<b_{II}<15{\degr}$ and 
$-60{\degr}<b_{II}<-30{\degr}$ (directions that we consider as likely dominated by field stars). Similar plots were produced 
for the stars observed at high resolution (see right panel of Fig.~\ref{fig:teffli}). The upper envelope for the Pleiades 
stars, adapted from the \cite{soderblometal1993} data, is also overplotted. In both panels, a spatial segregation of lithium 
strength can be observed, thus justifying the use of both high- and low-resolution $EW_{\rm Li}$. Practically all the 
stars located on the Orion region fall above the Pleiades upper envelope, as expected. The majority of these stars are indeed 
very young. Note that many stars, located in the region of the hypothetical Gould Belt, also have strong lithium 
absorption but tend to be closer to the Pleiades upper envelope. Finally, the majority of the lithium field stars 
fall closer to/or below the Pleiades upper envelope, and most of them seem to have an age similar to the Pleiades or older. 
Nevertheless, a few of these field stars have lithium strengths comparable to those of stars in Orion or 
the Gould Belt direction, and seem to be also very young.

\subsubsection{Lithium abundance} 
The lithium abundance ($\log n{\rm (Li)}$) was derived from the $EW_{\rm Li}^{\rm hr}$ and $T_{\rm eff}$ 
values and using the non-LTE curves-of-growth reported by \cite{pavlenkomagazzu1996}, assuming $\log g=4.5$. The main source 
of error in $\log n{\rm (Li)}$ is the uncertainty in $T_{\rm eff}$, which is about $\Delta T_{\rm eff}=150$ K. Taking this 
value and a mean error of about 10 m\AA~in $EW_{\rm Li}$ into account, we estimate a mean $\log n{\rm (Li)}$ error ranging 
from $\sim$0.22 dex for cool stars ($T_{\rm eff}\approx 3700$ K) down to $\sim$0.13 dex for warm stars ($T_{\rm eff}\approx 5800$ K). 
Moreover, the assumption of $\log g=4.5$ affects the lithium abundance determination, in the sense that the lower the surface 
gravity the higher the lithium abundance. In particular, the difference in $\log n{\rm (Li)}$ may rise to $\sim$0.1 dex, when 
considering stars with mean values of $EW_{\rm Li}=300$ m\AA~and $T_{\rm eff}=5000$ K and assuming $\Delta \log g=1.0$ dex. 
Hence, this means that our assumption of $\log g=4.5$ would eventually lead to underestimate the lithium abundance. 

\begin{figure*}	
\begin{center}
 \begin{tabular}{c}
\includegraphics[width=9.2cm]{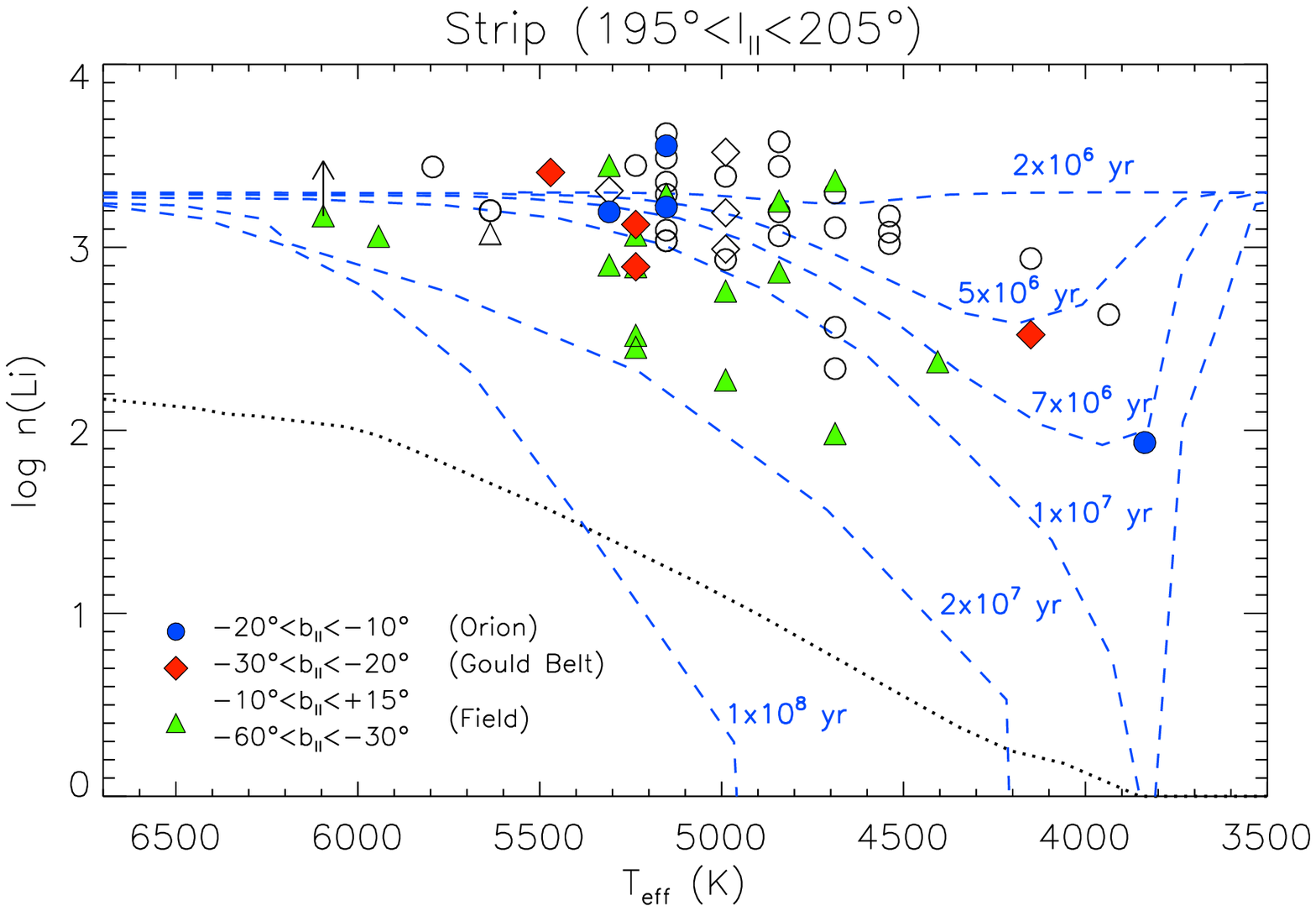}
\includegraphics[width=9.2cm]{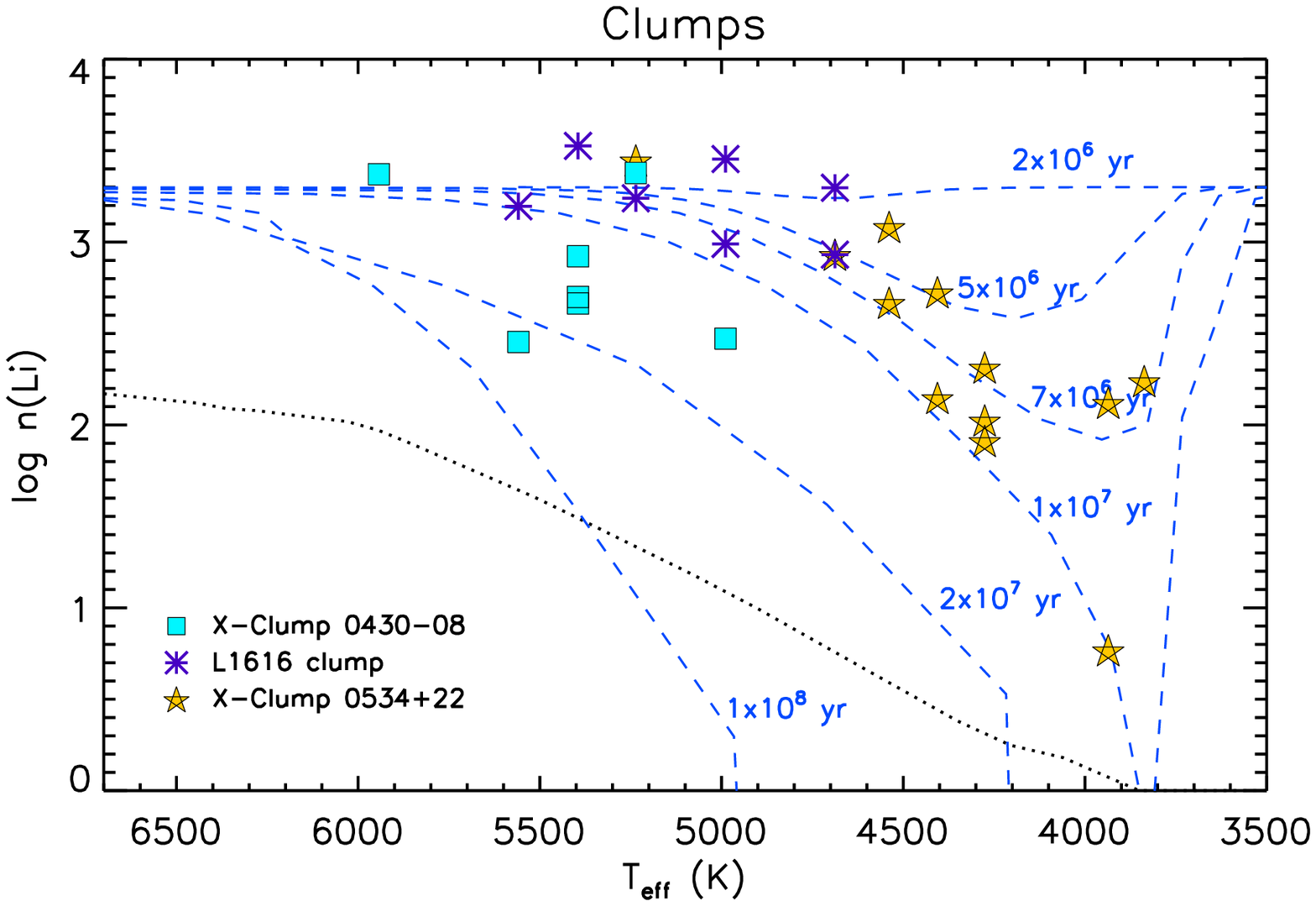}
 \end{tabular}
\caption{Lithium abundance versus effective temperature for stars observed at high resolution (filled symbols refer to our sample, 
while open symbols represent the \cite{alcalaetal2000} measurements). The ``lithium isochrones'' by \cite{dantonamazzitelli1997} 
in the $2\times10^6-1\times 10^8$ yr range are overlaid with dashed lines. Dotted lines represent the limiting detectable 
$\log n{\rm (Li)}$, as derived from \cite{pavlenkomagazzu1996} non-LTE curves-of-growth, assuming 10~m\AA~as mean $EW_{\rm Li}$ 
error and adding quadratically the mean contribution of the iron line at 6707.4~\AA~computed from the empirical relation 
obtained by \cite{soderblometal1993} between $B-V$ color and $EW_{\rm Fe}$ (\citealt{kenyonhartmann1995} tables were used to convert 
$B-V$ colors into $T_{\rm eff}$). {\it Left panel:} Stars  located along the strip crossing the GB in the range 
$195\degr<l_{II}<205\degr$ (see Fig.~\ref{fig:map_lowhigh}) and three bins of Galactic latitude. {\it Right panel:} Stars in 
known (L1616) and unknown (X-Clump 0430$-$08 and X-Clump 0534+22) groups.}
\label{fig:tefflognli}
 \end{center}
\end{figure*}

In Fig.~\ref{fig:tefflognli} (left panel), we show the lithium abundance as a function of the effective 
temperature for the stars on the strip, but coded in three bins of Galactic latitude, hence, according to their 
spatial location with respect to the Orion SFR and the Gould Belt. We also overplot the 
isochrones of lithium burning as calculated by \cite{dantonamazzitelli1997}. Three groups of stars can be identified
according to the lithium content and spatial location. First, stars showing high-lithium content with ages even 
younger than $\sim2-5\times10^6$ yr, mostly located on/or close to the Orion clouds; second, stars with lithium 
content consistent with ages $\sim5\times10^6-1\times 10^7$ yr, supposedly distributed on the Gould Belt, 
and third, stars with lithium indicating a wide range of ages, but located far off the Orion SFR or the GB.
 
In Fig.~\ref{fig:tefflognli} (right panel), we show the same plot, but for the three identified young aggregates,
respectively represented by three different symbols. The lithium content in the X-Clump 0430$-$08 corresponds to
an age of about $2\times10^6-2\times10^7$ yr, while for the L1616 group it indicates an age of $2-7\times10^6$ yr, 
consistent with the \cite{alcalaetal2004} and \cite{gandolfietal2008} findings. Finally, the lithium content 
of the stars in X-Clump 0534+22 indicates a relatively narrow age range of $2-10$ Myr, which is consistent 
with the age inferred from the HR diagram when adopting a distance of 140\,pc (see Sect.~\ref{sec:clumps}).

\subsubsection{Rotational and radial velocity measurements}
\label{sec:velocities} 

Stellar rotation may affect internal mixing, hence lithium depletion. A large spread in rotation 
rates may introduce a spread in lithium abundance, which is observed in young clusters 
(\citealt{balachandranetal2011}, and references therein). In order to investigate such Li behaviour 
in the stars of our sample, rotational ($v \sin i$) velocities were derived by using the same 
cross-correlation method as described in \cite{alcalaetal2000}.

\begin{figure*}	
\begin{center}
 \begin{tabular}{c}
\includegraphics[width=9.2cm]{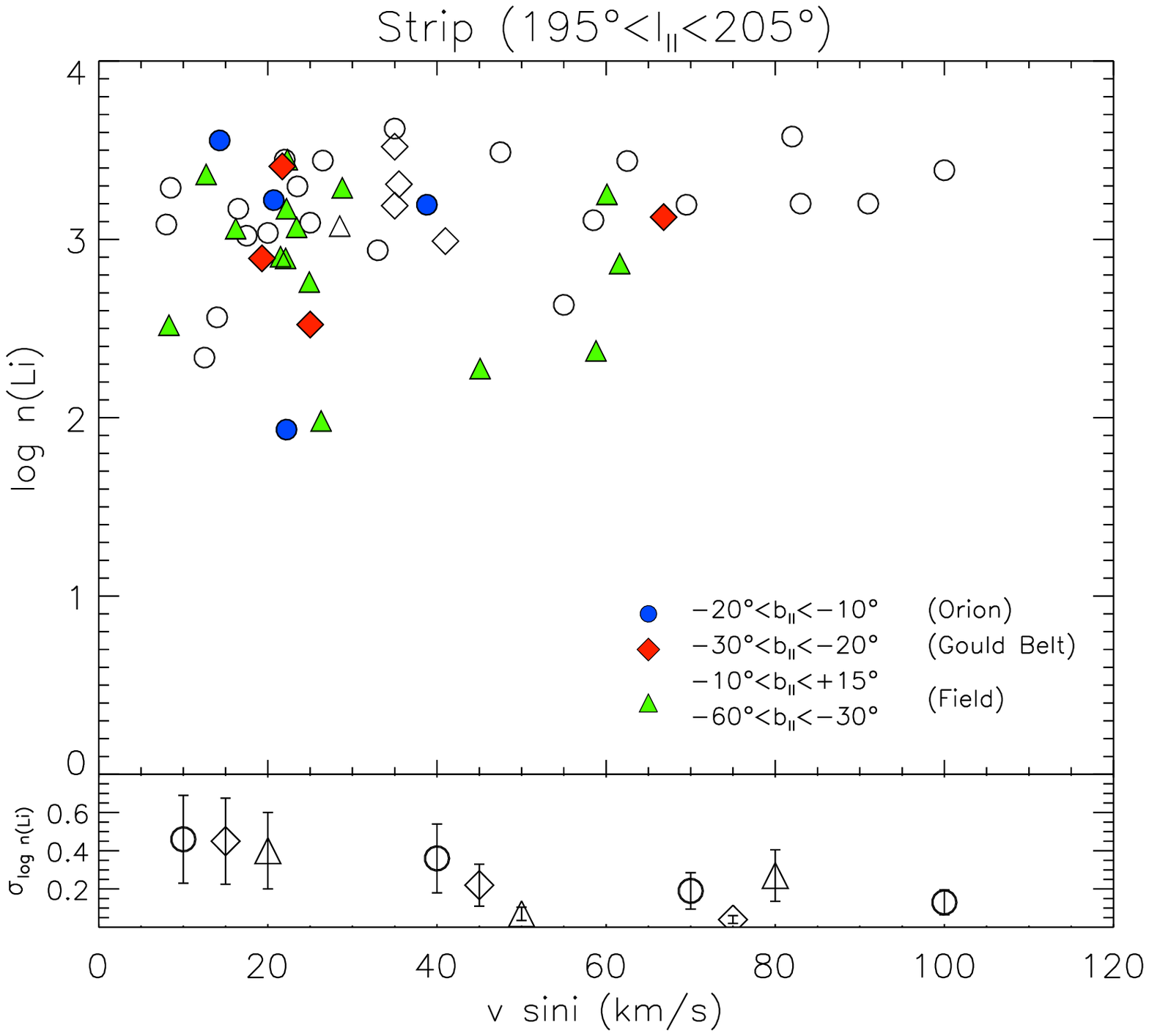}
\includegraphics[width=9.2cm]{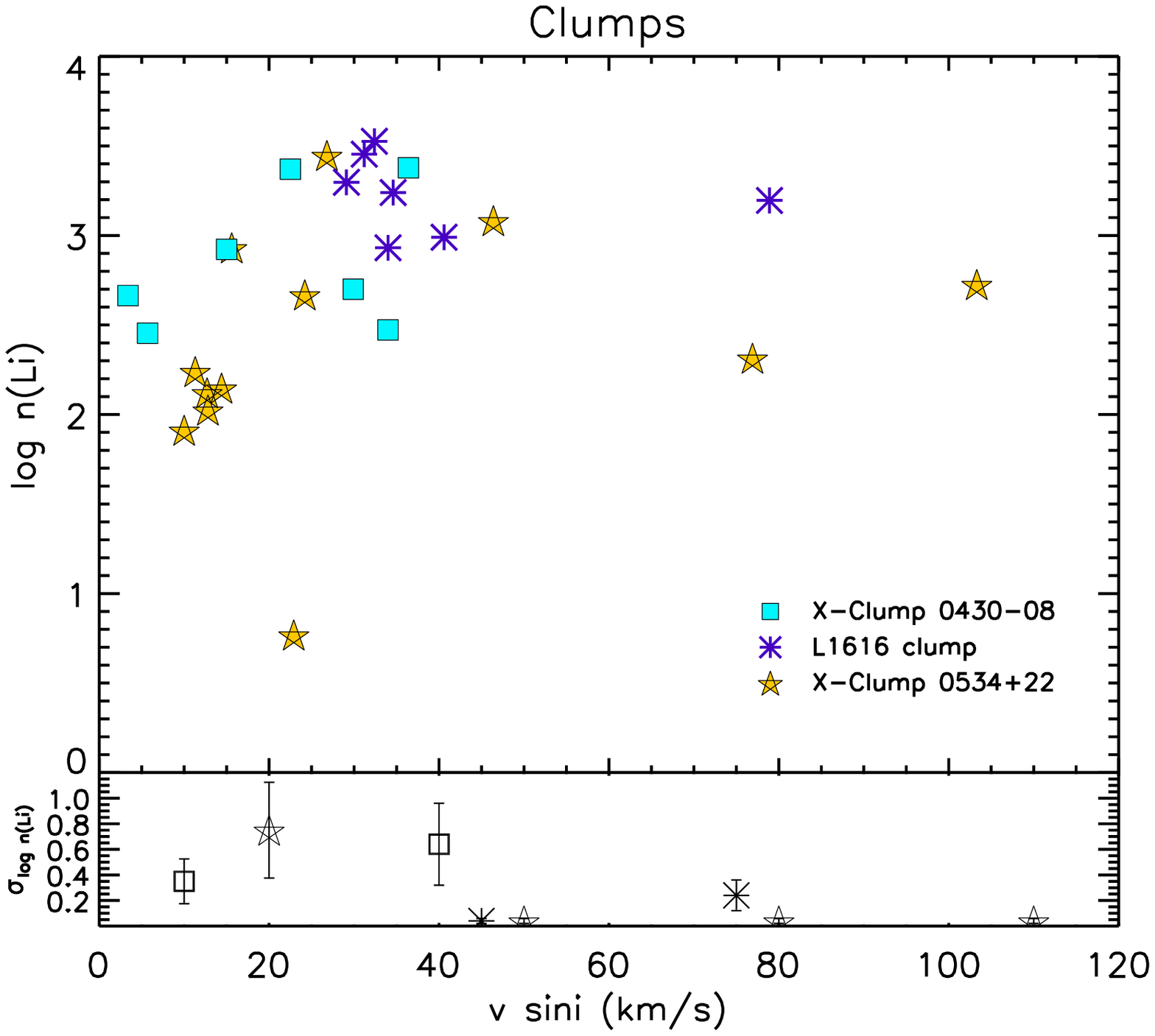}
 \end{tabular}
\caption{Lithium abundance versus $v \sin i$ for stars observed at high resolution. Symbols as in Fig.~\ref{fig:tefflognli}. 
Bottom panels show for each group the $\log n{\rm (Li)}$ dispersion as a function of $v \sin i$ at the following bins: 
$0-30$, $30-60$, $60-90$, $90-120$ km~s$^{-1}$. The values are shifted in $v \sin i$ to better visualize the symbols.}
\label{fig:vsinilognli}
 \end{center}
\end{figure*}

In Fig.~\ref{fig:vsinilognli} (left panel) we show the lithium abundance versus $v \sin i$ for the
on-strip stars, coded with three different symbols, indicating their spatial location as defined in the 
previous Section. Despite the low statistics, the typical $\log n{\rm (Li)}-v \sin i$ behaviour
observed in young clusters, i.e. a larger spread in $\log n{\rm (Li)}$ for lower $v \sin i$ values, is 
apparent for the stars projected in Orion and in the Gould Belt. In order to assign a confidence level 
to this trend, a one-side $2\times2$ Fisher's exact test\footnote{We used the following web calculator 
(\citealt{langsrudetal2007}): http://www.langsrud.com/fisher.htm.} was performed (\citealt{agresti1992}). 
For the test, we adopt 30 km s$^{-1}$ and 2.0 dex as dividing limits in $v \sin i$ and $\log n{\rm (Li)}$, 
respectively. We find a $p$-value of 0.54 as chance that random data would yield the trend, 
indicating a probability of correlation of 46\%. Hence, the low-number statistics prevents 
a rigorous demostration of the apparent trend shown in the plots. More measurements of 
$\log n{\rm (Li)}$ and $v \sin i$ for stars projected in Orion and the Gould Belt are needed
to firmly establish the preservation of Li content at high $v \sin i$ ($\gtsim 25$ km s$^{-1}$) 
values. The above $\log n{\rm (Li)}-v \sin i$ behaviour is even less evident, however,
for the stars flagged as ``field". These stars show a spread in $\log n{\rm (Li)}$ at all $v \sin i$ values. 
The difference between the three stellar groups is supported by the different dispersion in the
$\log n{\rm (Li)} - v \sin i$ diagram (see left-bottom panel of Fig.~\ref{fig:vsinilognli}). 

In Fig.~\ref{fig:vsinilognli} (right panel) the lithium abundance is plotted as a function of $v \sin i$ 
for the stars in the young aggregates. Similar conclusions 
can be achieved as for the young stars projected in Orion and the Gould Belt. The $\log n{\rm (Li)}-v \sin i$
behaviour in the X-Clump 0534+22 aggregate is enhanced by the star with the lowest lithium abundance 
and low $v \sin i$. This star, namely 2MASS~J06020094+1955290, shows basically the same activity level 
as the other targets (see $EW_{\rm H\alpha}$ in Table~\ref{tab:parameters} and $\log \frac{f_X}{f_V}$
values in the histograms of the top panel of Fig.~\ref{fig:strip}) and is indistinguishable from 
the other stars in the aggregate. 

\begin{figure*}	
\begin{center}
 \begin{tabular}{c}
\includegraphics[width=9.2cm]{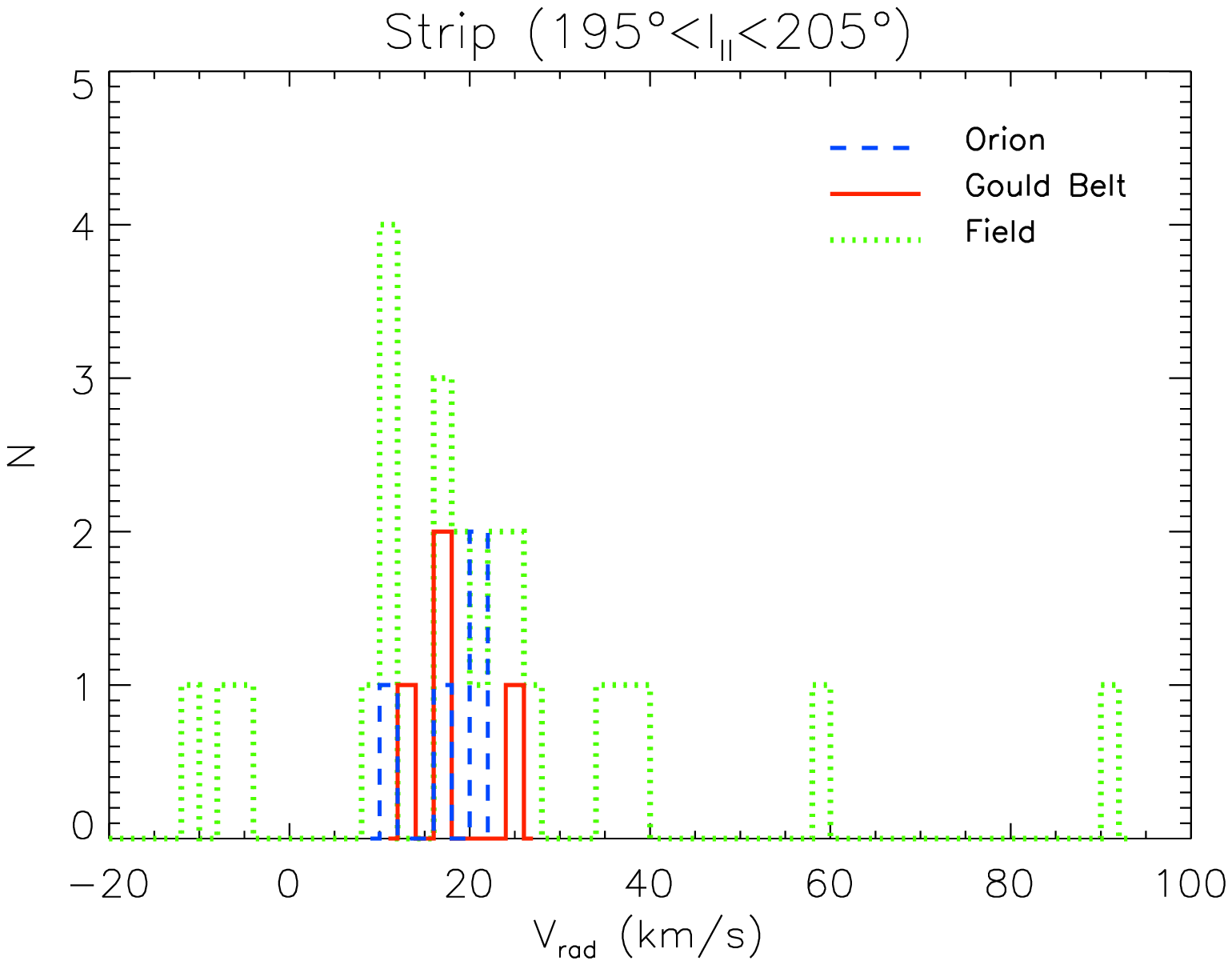}
\includegraphics[width=9.2cm]{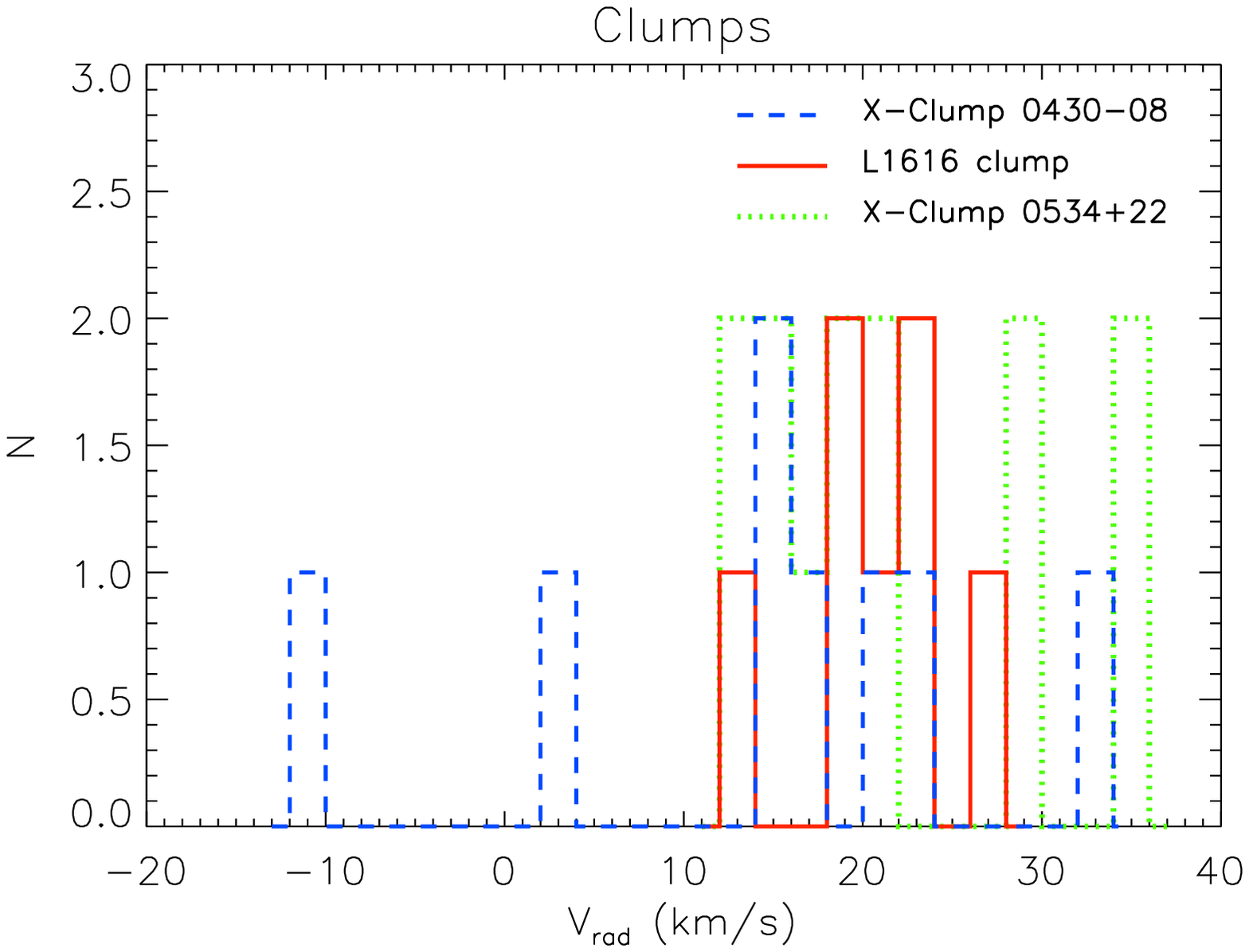}
 \end{tabular}
\caption{Radial velocity distribution for the on-strip stars ({\it left panel}) and in the young aggregates
 ({\it right panel}), respectively.} 
\label{fig:vrad_distr}
 \end{center}
\end{figure*}

Stellar radial velocities (RV) were also measured by means of cross-correlation analysis. In Fig.~\ref{fig:vrad_distr} 
(left panel) we show the distribution of RV for the on-strip stars, also divided in three bins of Galactic latitude as 
above. While the field stars show a wide RV range, the RV distribution of the stars projected on Orion and the GB is peaked 
at values of $\sim$18 km~s$^{-1}$, i.e. close to the tail at low RV of the Orion sub-associations (from 
19.7$\pm$1.7 km~s$^{-1}$ for 25~Ori to $24.87\pm2.74$ km~s$^{-1}$ for the ONC, to $\sim$30.1$\pm$1.9 km~s$^{-1}$ 
for OB1b; see, \citealt{bricenoetal2007,biazzoetal2009}) or to the Taurus-Auriga distribution (16.03$\pm$6.43 km~s$^{-1}$; 
\citealt{bertoutgenova2006}). 

In  Fig.~\ref{fig:vrad_distr} (right panel) the RV distribution of the stars in the young aggregates is shown.
With the exception of two stars in the X-Clump 0430$-$08, likely spectroscopic binaries, the RV 
of these groups is in the range 10--35 km~s$^{-1}$, fairly consistent with Orion or Taurus.
It is worth mentioning that the RV distribution of widely distributed young stars in Orion shows 
a double peak (\citealt{alcalaetal2000}), which can be explained as due to objects
associated with different kinematical groups, likely located at different distances.

\subsubsection{Iron abundance} 
Metallicity measurements were obtained following the prescriptions by \cite{biazzoetal2011a,biazzoetal2011b} 
and the 2010 version of the MOOG\footnote{http://www.as.utexas.edu/$\sim$chris/moog.html} code (\citealt{sneden1973}). 

After a screening of the sample for the selection of suitable stars (G0-K7 stars with $v\sin i\ltsim20$ km s$^{-1}$
and no evidence of multiplicity), a total of 11 stars in our sample, plus 8 stars from \cite{alcalaetal2000} were 
analyzed for iron abundance measurements. In Table~\ref{tab:iron_abun} we list the final results, together with effective 
temperature, surface gravity, and microturbulence (the number of lines used is also given in Columns 6 and 8). An initial 
temperature value was set using the ARES\footnote{http://www.astro.up.pt/$\sim$sousasag/ares/} automatic code 
(\citealt{sousaetal2007}); initial microturbulence was set to $\xi=1.5$ km\,s$^{-1}$, and initial gravity to 
$\log g=4.0$. The effective temperatures derived using this method and from spectral types (Sect.~\ref{sec:sptype_lithium_halpha}) 
agree within $\sim$200\,K (i.e. $\sim$1.5 spectral sub-class) on the average. For the stars observed with both FOCES and FEROS 
spectrographs, the values of the stellar parameters are in close agreement. It is worth noticing that the target 
2MASS~J05214684+2400444 was also analyzed by Santos et al. (2008; their 1RXSJ052146.7), who derived stellar parameters 
($T_{\rm eff}=4921\pm59$ K, $\log g=4.05\pm0.29$, $\xi=1.92\pm0.07$ km s$^{-1}$, $v \sin i=13$ km s$^{-1}$) and iron 
abundance ([Fe/H]$=-0.07\pm0.07$) in good agreement with our determinations, thus excluding any significant systematic 
error due to different datasets (see \citealt{biazzoetal2011b}).

\begin{figure}  
\begin{center}
 \begin{tabular}{c}
\hspace{-0.7cm}
\includegraphics[width=9cm]{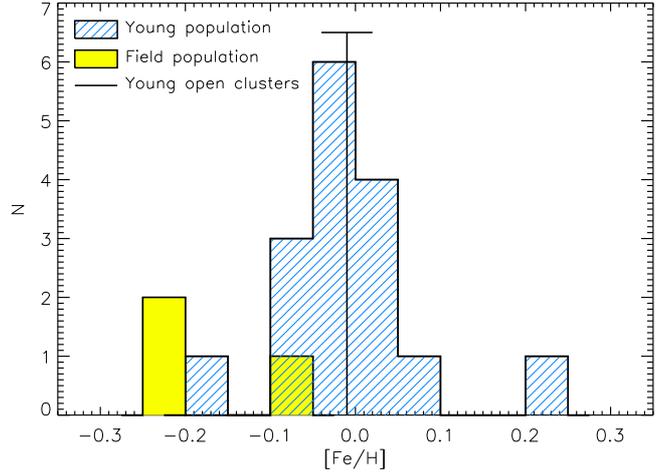}
 \end{tabular}
\caption{[Fe/H] distribution for stars observed with the FOCES spectrograph and representing the young population (dashed 
histogram) and the field population (filled histogram). The bar represents the mean value of [Fe/H]=$-0.01\pm0.03$ obtained 
for open clusters younger than 150 Myr within 500 pc from the Sun 
(see \citealt{biazzoetal2011a}, and references therein).} 
\label{fig:fe_distr}
 \end{center}
\end{figure}

In Figs.~\ref{fig:fe_distr} and \ref{fig:map_fe} we show the distribution of iron abundance for the stars with strong 
lithium absorption (16 stars representing the young population) and those without (3 stars representing the 
old field population). Two results are noticeable: $i)$ the stars with high lithium content show a distribution with a 
mean [Fe/H]\footnote{[Fe/H]=$\log n^\star{\rm (\ion{Fe}{i})}-\log n^\odot{\rm (\ion{Fe}{i})}$}$=-0.02\pm0.09$, 
consistent with the value of $-0.01\pm0.03$ for open clusters younger than 150 Myr and within 500 pc from the Sun (see 
\citealt{biazzoetal2011a} for details); $ii)$ stars with no lithium absorption line show [Fe/H] values which 
are within the metallicity distribution of field stars in the solar neighborhood ([Fe/H]=$-0.10\pm0.24$; 
\citealt{santosetal2008}). This would imply that the young stars in the Orion vicinity have solar metallicity, 
consistent with the distribution of the Galactic thin disk in the solar neighborhood (\citealt{biazzoetal2011b}).

\begin{table*}
\caption{Spectroscopic stellar parameters and iron abundance derived with MOOG for low-rotating stars observed at high resolution. } 
\label{tab:iron_abun}
\begin{center}
\tiny
\begin{tabular}{cccccccc}
\hline
\hline
Star & $T_{\rm eff}$ & $\log g$ & $\xi$ &[\ion{Fe}{i}/H]$^*$& \# lines &[\ion{Fe}{ii}/H]$^{**}$& \# lines \\
  & (K) & (dex) & (km/s) & (dex) &   & (dex) &  \\
\hline
~\\
\multicolumn{8}{c}{{\sc ~~~~~~~~~~~~~~~~~~~~~~~~~~~~~~~~X-Clump 0534+22~~~~~~~~~~~~~~~~~~~~~~~~~~~~~}}\\
\hline
2MASS~J05263833+2231546 & 4750 & 4.4 & 2.0 & $-0.08\pm0.15$ & 43 & $-0.08\pm0.15$ & 3 \\
2MASS~J05263826+2231434 & 4750 & 4.0 & 1.9 & $-0.07\pm0.12$ & 42 & $-0.06\pm0.11$ & 5 \\
2MASS~J05214684+2400444 & 5000 & 4.3 & 2.2 & $-0.04\pm0.11$ & 44 & $-0.02\pm0.06$ & 4 \\
\hline
~\\
\multicolumn{8}{c}{{\sc ~~~~~~~~~~~~~~~~~~~~~~~~~~~~~~X-Clump 0430$-$08~~~~~~~~~~~~~~~~~~~~~~~~~~~~~}}\\
\hline
2MASS~J04431640$-$0937052 & 6000 & 4.6 & 1.4 & $+0.22\pm0.08$ & 66 & $+0.22\pm0.10$ & 9 \\
{\it "}      & {\it 6050} & {\it 4.6} & {\it 1.4} & {\it +0.26$\pm$0.08} & {\it 78} & {\it +0.22$\pm$0.10} & {\it 9 }\\
2MASS~J04443859$-$0724378 & 5900 & 4.4 & 1.5 & $+0.03\pm0.09$ & 64 & $+0.03\pm0.08$ & 11 \\
\hline
~\\
\multicolumn{8}{c}{{\sc ~~~~~~~~~~~~~~~~~~~~~~~~~~~~~~~~Strip~~~~~~~~~~~~~~~~~~~~~~~~~~~~~}}\\
\hline
2MASS~J03494386$-$1353108 & 5400 & 4.5 & 1.4 & $+0.00\pm0.09$ & 64 & $-0.01\pm0.12$ & 9 \\
     {\it "}              & {\it 5500} & {\it 4.6} & {\it 1.6} & {\it +0.05$\pm$0.09} & {\it 70} & {\it +0.03$\pm$0.13} & {\it 10}\\
2MASS~J05274490+0313161   & 5000 & 4.3 & 1.7 & $-0.02\pm0.17$ & 51 & $-0.01\pm0.23$ & 3 \\
2MASS~J06134773+0846022   & 5050 & 3.6 & 1.6 & $-0.25\pm0.13$ & 59 & $-0.24\pm0.13$ & 9 \\
2MASS~J06203205+1331125   & 5350 & 4.4 & 0.3 & $-0.23\pm0.17$ & 44 & $-0.24\pm0.25$ & 4 \\
2MASS~J06350191+1211359   & 6100 & 4.3 & 1.5 & $+0.01\pm0.13$ & 45 & $+0.00\pm0.19$ & 9 \\
2MASS~J06513955+1828080   & 5100 & 4.7 & 1.5 & $-0.07\pm0.09$ & 60 & $-0.07\pm0.13$ & 5 \\
\hline
~\\
\multicolumn{8}{c}{{\sc ~~~~~~~~~~~~~~~~~~~~~~Alcal\'a et al. (2000) Sample~~~~~~~~~~~~~~~~~~~~~}}\\
\hline
RX~J0515.6$-$0930 & 5650 & 4.5 & 2.3 & $-0.04\pm0.12$ & 41 & $-0.06\pm0.23$ & 4 \\
RX~J0517.9$-$0708 & 5100 & 4.2 & 1.8 & $+0.06\pm0.13$ & 47 & $+0.05\pm0.14$ & 4 \\
RX~J0531.6$-$0326 & 5250 & 4.3 & 1.7 & $-0.02\pm0.09$ & 53 & $-0.03\pm0.08$ & 5 \\
RX~J0538.9$-$0624 & 5550 & 4.4 & 1.3 & $+0.01\pm0.11$ & 58 & $+0.02\pm0.20$ & 10 \\
RX~J0518.3+0829   & 5300 & 4.2 & 2.0 & $-0.04\pm0.11$ & 21 & $-0.02\pm0.15$ & 3 \\
RX~J0510.1$-$0427 & 4950 & 4.6 & 1.9 & $-0.10\pm0.12$ & 34 & $-0.00\pm0.24$ & 2 \\
RX~J0520.0+0612   & 4750 & 4.1 & 2.2 & $-0.17\pm0.12$ & 45 & $-0.17\pm0.30$ & 5 \\
RX~J0520.5+0616   & 4900 & 4.1 & 2.4 & $-0.08\pm0.17$ & 29 & $-0.07\pm0.04$ & 3 \\
\hline
\end{tabular}
\end{center}
\footnotesize{Note: The two stars observed with FEROS are indicated in italics. \\
$^{*}$ The iron abundances are relative to the Sun. Adopting $T_{\rm eff}^\odot=5770$ K, 
$\log g^\odot=4.44$, and $\xi^\odot=1.1$ km s$^{-1}$, we obtain $\log n^\odot{\rm (\ion{Fe}{i})}=7.53\pm0.04$ 
and $\log n^\odot{\rm (\ion{Fe}{ii})}=7.53\pm0.06$ from a FOCES spectrum, and 
$\log n^\odot{\rm (\ion{Fe}{i})}=7.50\pm0.05$ and $\log n^\odot{\rm (\ion{Fe}{ii})}=7.50\pm0.06$ from 
a FEROS spectrum.\\
$^{**}$ The listed errors are the internal ones in $EW$, represented by the standard deviation on the mean 
abundance determined from all lines. The other source of internal error includes uncertainties in 
stellar parameters (\citealt{biazzoetal2011a}). Taking into account typical errors in 
$T_{\rm eff}$ ($\sim$70 K), $\log g$ ($\sim$0.2 dex), and $\xi$ ($\sim$0.2 dex), we derive an error of 
$\sim$0.05 dex in [Fe/H] due to uncertainties on stellar parameters.}
\end{table*}
\normalsize

\section{Discussion}
\label{sec:discussion}
\subsection{RASS and Galactic Models}
How many young, X-ray active stars are actually expected inside the strip? A comparison of the {\sl number counts} with predictions 
from Galactic models provides the basis for a quantitative analysis of source excesses (or deficits) in order to understand their 
origin. Because of the strong dependence of stellar X-ray emission on age, an X-ray view of the sky preferentially reveals
young stars (ages $\le 100$~Myr), in contrast to optical star counts which only loosely constrain the stellar population for 
ages $\le 10^9$~yr. A Galactic X-ray star count modeling starts  adopting a Galactic model, including assumptions about the 
spatial and temporal evolution of the star formation rate and the initial mass function, and uses characteristic X-ray luminosity 
functions attributed to the different stellar populations. Such models are able to predict the number of stars per square degree 
$N(>S)$ with X-ray flux $>S$, taking into account the dependence on Galactic latitude, spectral type, and stellar age. An elaborate 
Galactic model, including kinematics, is the {\it evolution synthesis population model} developed at Besan\c{c}on 
(\citealt{robincreze1986}), which computes the density and the distribution of stars as a function of the observing direction, 
age, spectral type, and distance. Our X-ray synthetic model is based on the Besan\c{c}on optical model and has been first 
applied to the analysis of the RASS stellar population by \cite{guilloutetal1996}. \cite{motchetal1997} successfully used 
this model in a low Galactic latitude RASS area in Cygnus and found a good agreement between observations and predicted number counts 
using the `canonical' assumption of a uniform and continuous star formation history in the solar vicinity. We note however that, 
following the publication of {\it Hipparcos} results, the stellar density in the solar neighborhood was revised (lowered) in 
the Besan\c{c}on model thus propagating in the X-ray population model predictions. The apparent disagreement between observed 
and predicted number count (by $\sim$20\%) can in fact be explained by the population of old close binaries (RS CVn-like systems), 
as suggested by \cite{favataetal1988} and \cite{sciortinoetal1995}. RS CVn systems for which the high magnetic activity level 
results from the synchronization of rotational and orbital periods can mimic young active stars and contaminate the young star 
population detected in soft X-ray survey (\citealt{frascaetal2006, guilloutetal2009}). The eight stars with 
H$\alpha$ emission, but no Li absorption identified by us (c.f. Section~\ref{sec:sptype_lithium_halpha}) may represent this type 
of objects.

\begin{figure}	
\begin{center}
 \begin{tabular}{c}
\includegraphics[width=9cm]{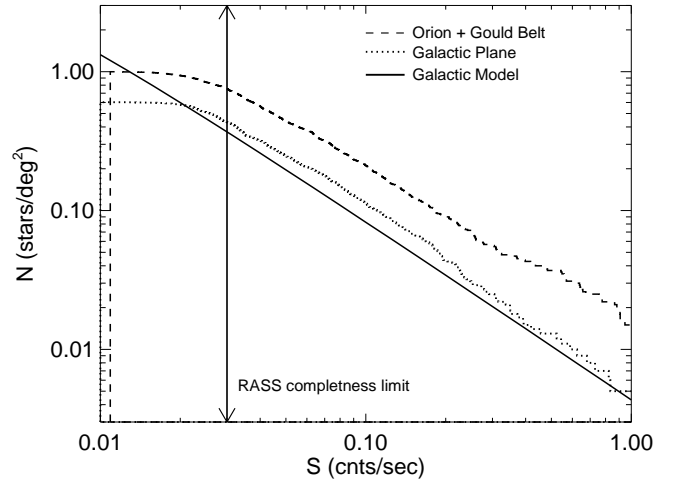}
 \end{tabular}
\vspace{-0.5cm}
\caption{Cumulative $\log N(>S)-\log S$ distributions for RASS data in the Orion vicinity. A field centered on the Galactic Plane 
is shown with a dotted line. The dashed line includes Orion and a section of the Gould Belt. The solid line corresponds to the prediction 
of our X-ray models at $l_{II}=200\degr$ and zero Galactic latitude. The RASS completeness limit of 0.03~cnts/sec is marked with the 
double arrow.
} 
\label{fig:lnls}
 \end{center}
\end{figure}

In Fig.~\ref{fig:lnls}, we compare the RASS stellar counterparts (taken from the Guide Star Catalogue) in our field with the current 
X-ray Galactic model predictions using a cumulative distribution function $\log N(>S)-\log S$. We select two fields: one centered 
on the Galactic Plane ($b_{II}=0\degr \pm 10\degr$, $190\degr<l_{II}<220\degr$), and the other, southern, includes Orion and 
a section of the Gould Belt ($b_{II}=-20\degr \pm 10\degr$, $190\degr<l_{II}<220\degr$). The deviation from a power-law 
function for both data distributions at a count rate of $\approx$0.03~cnts/sec is related to the completeness limit of the RASS at this 
value. Comparing these two distributions, it is noticeable how the source density in the area containing Orion exceeds that of the 
Galactic Plane by about 30$-$60\% (at the largest count rates even by a factor of two). Considering that the population of old active 
binaries is not yet taken into account in our X-ray model, the predictions are in close agreement with the RASS data around 
the Galactic Plane and with optical counterparts from the Guide Star Catalogue. For the Galactic Plane, this model predicts 
a surface density of 0.37~stars/deg$^2$ at the RASS completeness limit (see Table~\ref{tab:surface_density}). The source excess 
in the Orion field can be attributed to the presence of additional, probably younger, X-ray active stars due to recent, more 
localized star formation which is not included in the Galactic model. In fact, the difference in surface density between the 
regions in Orion + Gould Belt and the Gould Belt alone is 0.14$\pm$0.04 stars/deg$^2$ for the RASS stellar counterparts and 
of 0.21$\pm$0.07 stars/deg$^2$ for the stars with lithium detection, i.e. a significant number of 
young stars in the general direction of Orion, not necesarily originated in the star formation complex, is evident
(see Table~\ref{tab:surface_density}, where the errors were computed from Poisson statistics). We note that our results 
are not influenced by the density of active stars: from our spectroscopic identifications, we observed 61 stars in 
the strip out of 198 young star candidates resulting from the \cite{sterziketal1995} selection criteria. 
Eight are likely active stars because of their H$\alpha$ emission but no Li detection
(see Sect.~\ref{sec:sptype_lithium_halpha}). Therefore, inside the 750 deg$^2$ strip area, we estimate a surface 
density of only 0.03 stars/deg$^2$ as due to active stars.

Encouraged by the success of our X-ray star count modeling in reproducing the background counts and in revealing 
the excess of sources associated with Orion, we then performed a more detailed analysis of the RASS sources located 
in the strip shown in Fig.~\ref{fig:map}, including the available information on spectroscopic identification and 
lithium abundance. Our goal is to intercompare average source densities of RASS-selected young stars and to constrain 
the age distributions in different parts of the strip. Therefore, we divided the strip in three subareas, one is a 
200~deg$^2$ large field centered on the Galactic Plane and with $-10\degr<b_{II}<+10\degr$, the adjacent 100~deg$^2$ 
area between $-20\degr<b_{II}<-10\degr$ contains the northern parts of the Orion Complex, and the third 100~deg$^2$ 
area between $-30\degr<b_{II}<-20\degr$ is formally unrelated to the Orion molecular clouds but contains a significant 
part of the Gould Belt in that direction. 

\begin{figure}	
\begin{center}
 \begin{tabular}{c}
\includegraphics[width=9cm]{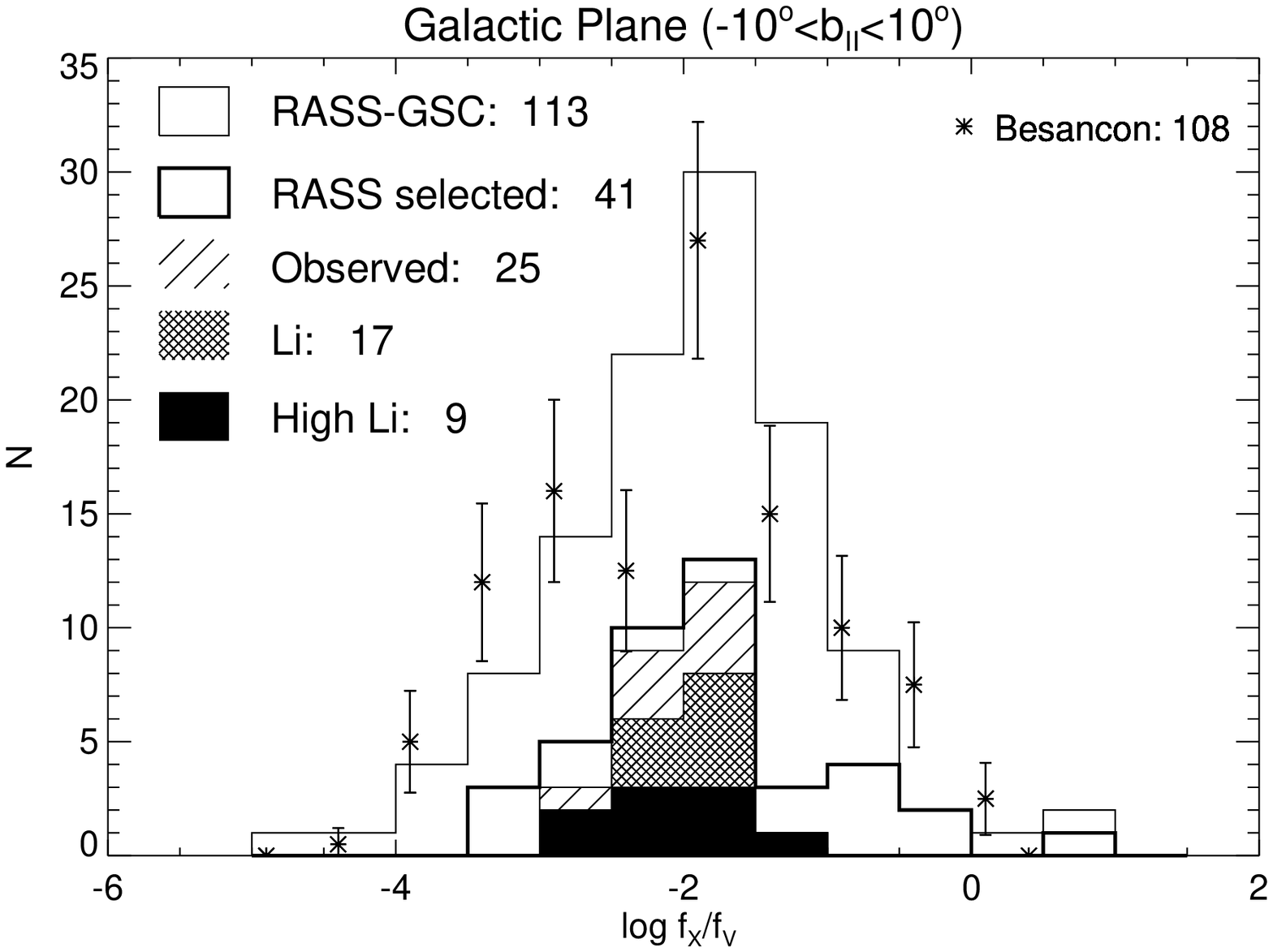}\\
\includegraphics[width=9cm]{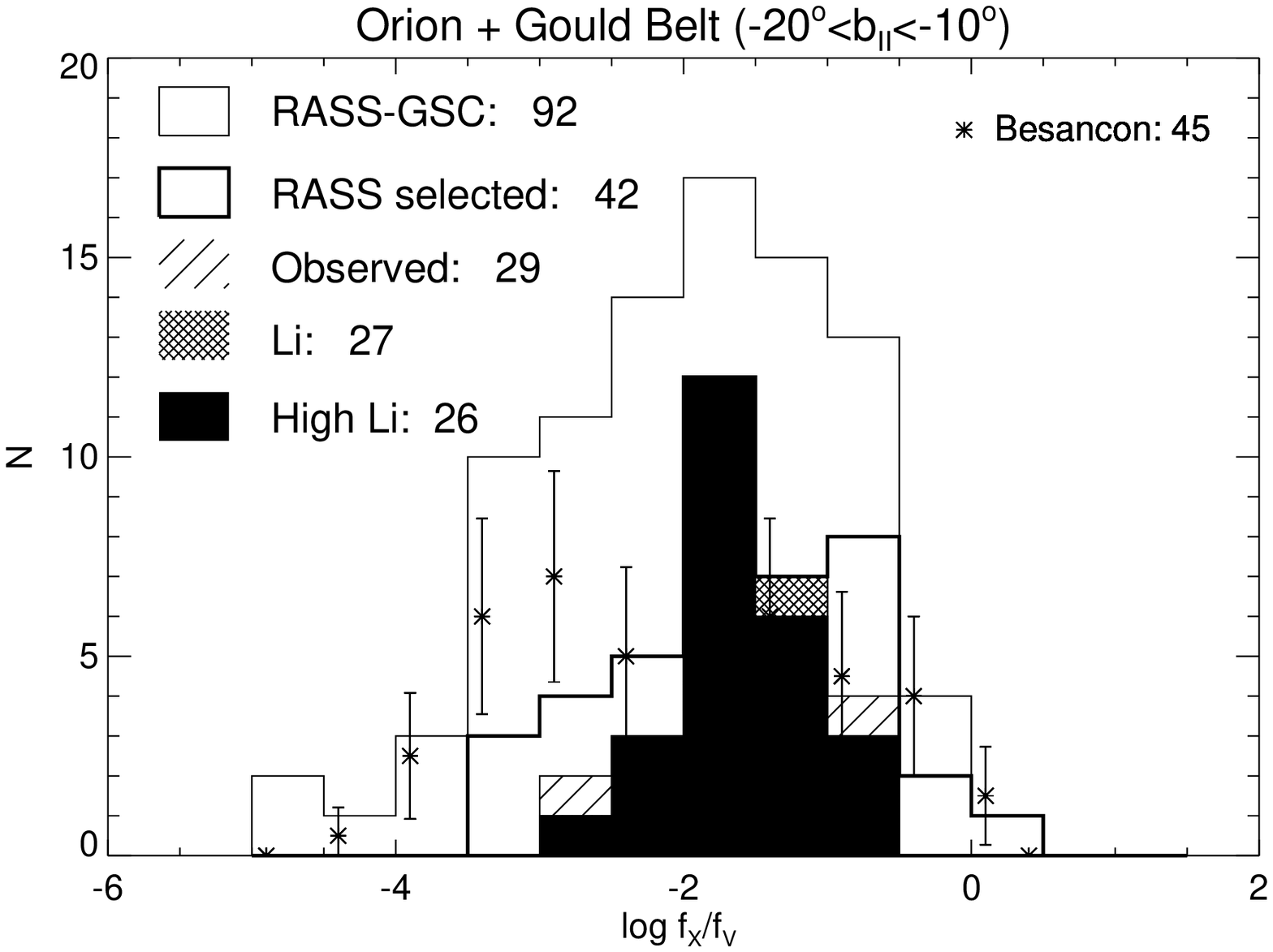}\\
\includegraphics[width=9cm]{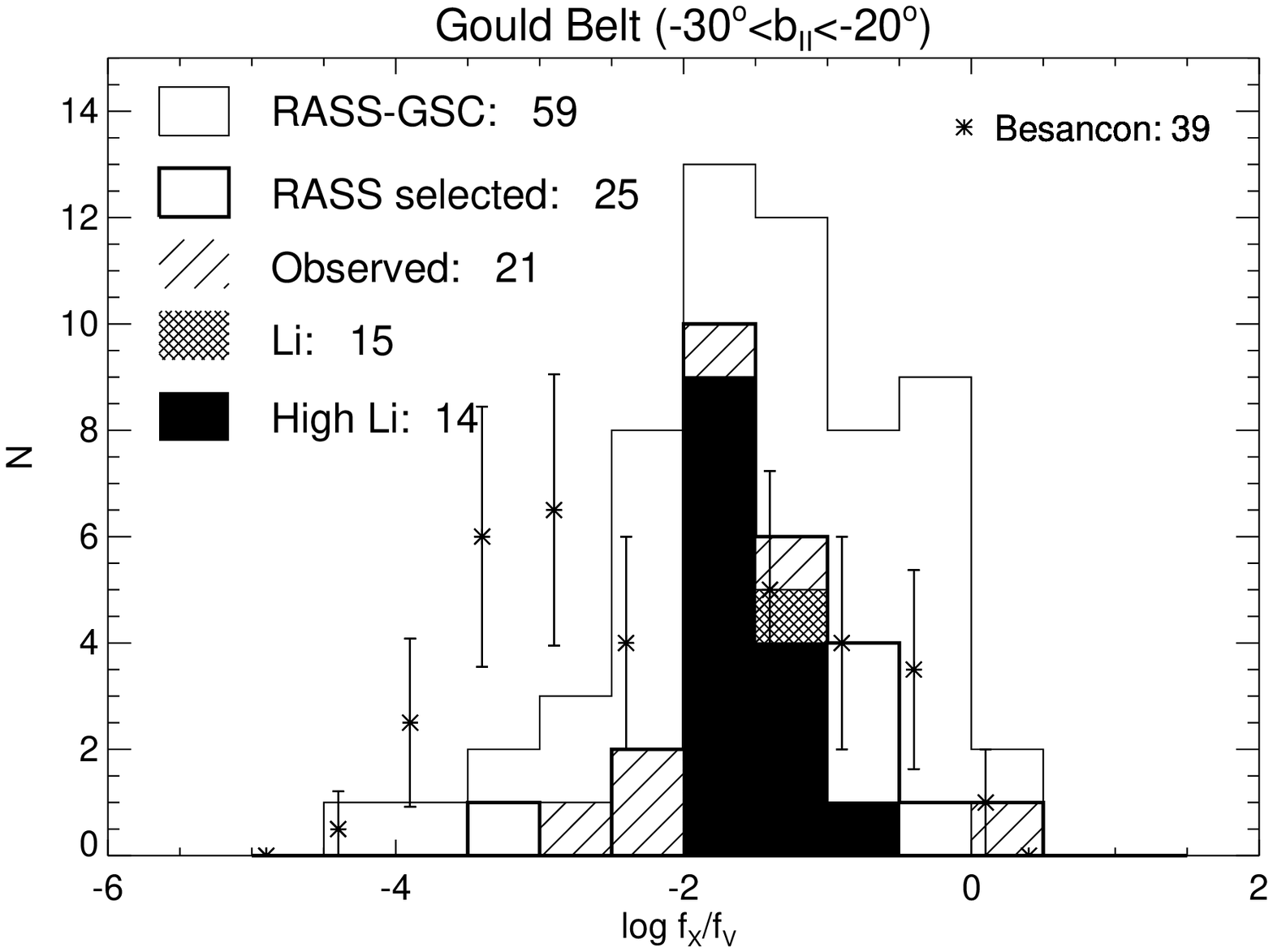}
 \end{tabular}
\caption{Detailed comparison of three fields in the strip perpendicular to the Galactic Plane between $l_{II}=195\degr$ and 
$l_{II}=205\degr$, as shown in Fig.~\ref{fig:ccd_2mass}. The three areas include: the Galactic Plane, the Orion SFRs, and the Gould 
Belt, respectively. Histograms are cumulative, not stacked, i.e., each histogram includes the stars in all subcategory histograms.}
\label{fig:strip}
 \end{center}
\end{figure}

In Fig.~\ref{fig:strip} we compare the number of RASS sources which have stellar counterparts from the Guide Star Catalogue 
and count rates $\ge 0.03$~cnts/sec (as indicated by the thin-lined histogram) with the numbers of such sources predicted 
by the Galactic model (indicated by star symbols with statistical error bars). The histograms show the X-ray to optical flux ratio 
distributions. The thick-lined histogram indicates the RASS sources that have been selected as young star 
candidates according to the \cite{sterziketal1995} criteria. A subsample of those (hatched histogram) were 
observed spectroscopically and classified according to lithium absorption strength. The dark grey histogram denotes 
objects where lithium absorption has been found, and the solid histogram refers to high-lithium stars. 
We note that the `identified' subsample is by now only complete to 40$-$70\% depending on the area. We can draw the 
following main conclusions:
\begin{enumerate}
\item The total number and the flux ratio distribution of the RASS sources are in good agreement with the Galactic model for the 
Galactic Plane field. High-lithium sources represent about 1/3 of the observed sources. Extrapolating to all RASS-selected 
young star candidates, we expect a total of about 15 high-lithium sources in this 200~deg$^2$ large Galactic Plane field. 
The Besan\c{c}on model predicts 72 X-ray sources with ages lower than 150~Myr in this area. The high-lithium sources are 
expected to be the youngest within our sample, and we can roughly estimate their characteristic age to be around 
15/72~$\cdot$~150~Myr=30~Myr (assuming continuous star formation). This age is consistent with their spectroscopic signature.
\item A large RASS source excess ($\sim$ 100\%) is present in the Orion field. The fraction of high-lithium stars among 
identified sources is now around 9/10. In this 100~deg$^2$ field, we expect to find a total of 38 high-lithium stars once 
all young candidates have been observed. 
\item A large RASS source excess (50\%) is also present in the field that includes the Gould Belt, preferentially at flux ratios 
$\log \frac{f_X}{f_V}>-2$. The fraction of high-lithium stars among observed sources is about 2/3. We expect eventually to 
find a total of about 17 high-lithium stars in the RASS-selected subsample. 
\end{enumerate}
These extrapolations to the total number of expected high-lithium sources should be reliable enough, since the spectroscopic 
identifications were done on statistically representative subsamples of RASS-selected young star candidates. It is also 
possible that some of the high-lithium stars present among RASS sources have been missed by the \cite{sterziketal1995} 
selection criteria. Hence, the extrapolated number of high-lithium X-ray sources in the respective areas is a lower limit. 

The results of this Section are summarized in Table~\ref{tab:surface_density}, where we report the estimated surface 
density of RASS sources and of high-lithium RASS-selected sources in the three investigated sky areas. While the Galactic 
Plane source densities are in good agreement with the Galactic X-ray star count model, the source density excesses in the other 
two directions indicate the presence of additional populations of high-lithium stars younger than 150~Myr. 

\begin{table*}
\caption{Surface densities (in stars/deg$^2$) of the RASS sources and high-lithium stars compared 
with the densities as predicted by the X-ray Galactic model for the appropriate Galactic latitude.} 
\label{tab:surface_density}
\begin{center}
\begin{tabular}{lccc}
\hline
\hline
	 	       & Galactic Plane & Orion+Gould Belt & Gould Belt \\ 
\hline
RASS ($>$0.03cnts/sec)      & 0.43$\pm$0.02 & 0.76$\pm$0.03 & 0.62$\pm$0.02 \\
Besan\c{c}on Galactic model & 0.37        & 0.37	     & 0.31 \\
Lithium sources	    & 0.14$\pm$0.02 & 0.39$\pm$0.06 & 0.18$\pm$0.04 \\
\hline
\end{tabular}
\end{center}
\end{table*}

\subsection{The X-ray clumps 0534+22 and 0430$-$08: two new young stellar aggregates?}
\label{sec:clumps} 
\begin{figure} 
\begin{center}
 \begin{tabular}{c}
\includegraphics[width=9cm]{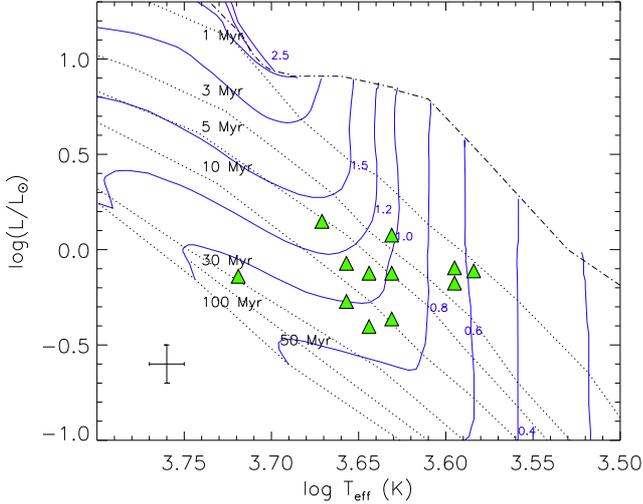}
 \end{tabular}
\caption{HR diagram for the stars in the X-Clump 0534+22. The pre-main sequence evolutionary 
tracks of \cite{pallastahler1999}, together with their isochrones and birthline are shown by solid, dotted, and dash-dotted lines, 
respectively. Labels indicating the track masses and the isochronal ages (from 1 to 100 Myr) are also overlaid. 
}
\label{fig:hr_diagram}
 \end{center}
\end{figure}

These two groupings were first recognized as surface density enhancements of X-ray emitting young star 
candidates (see Fig.~\ref{fig:map}), and then classified as young star clumps based on the spectroscopic follow-up. 
In particular, from lithium abundance measurements, the stars in the X-Clump 0534+22 have an age of $\sim5-10$ Myr, 
while those in the X-Clump 0430$-$08 of $\sim2-20$ Myr. Inspection of recent CO maps (see Fig.~\ref{fig:map_lowhigh}) 
seems to indicate that the stars in the X-Clump 0534+22 follow the spatial location of some of the Taurus-Auriga dark 
nebulae (namely L1545, L1548, and L1549; \citealt{kenyonetal2008}). Thus, the X-Clump 0534+22 might be an extension 
of the Taurus-Auriga association towards the Galactic Plane, alghough the average RV ($\sim22\pm8$ km s$^{-1}$) of 
this clump is higher than the mean RV for Taurus, but within the quoted uncertainty. Therefore, it deserves further 
kinematic studies. We attempt to estimate the stellar luminosities of the X-Clump 0534+22 stars by assuming the distance 
to Taurus ($d\sim140$ pc; \citealt{kenyonetal1994}), and adopting the {\it Tycho} magnitudes ($V_{\rm Tycho}$; 
\citealt{Egretetal1992}), a solar bolometric magnitude of $M^{\sun}_{\rm bol}=4.64$ (\citealt{cox2000}), and the 
temperatures listed in Table~\ref{tab:parameters}. The resulting HR diagram is shown in Fig.~\ref{fig:hr_diagram}.
The error in $\log L/L_{\sun}$ is estimated considering an uncertainty of 0.1 mag in $V_{\rm Tycho}$. The range of 
ages $\sim3-30$ Myr for the stars in the X-Clump 0534+22 appears consistent with the estimates from the lithium abundance 
(see Fig.~\ref{fig:tefflognli}). Adopting instead the Orion distance ($\sim400$ pc; \citealt{bally2008}) would place the 
stars close to/or above the birthline in the HR diagram, in contrast with their lithium content and lack of IR excess.

\subsection{The variety of populations in the Orion vicinity}
Based on the morphology and surface density distribution of the X-ray selected young star candidates, the 
spectroscopic analysis, and the comparison of RASS number counts with Galactic model predictions, we argue that 
the analyzed widespread stellar sample consists of a mixture of {\sl three distinct populations}:
\begin{enumerate}
\item The {\it clustered population} comprises the dense regions ($\gtsim 5-10$ stars/deg$^2$) associated with sites of 
active or recent star formation (e.g., OB1a, OB1b, OB1c, $\lambda$~Ori, L1616). These clusters contain the highest fraction 
of lithium-rich stars. The number counts result considerably in excess when compared to the Galactic model predictions. 
The stars in these clusters are, on average, among the youngest in our sample and their small spatial dispersion allows us 
to locate their birthplace. The RASS limit in X-ray luminosity at $M\sim0.8M_\odot$ is $\sim \log L_{\rm X}=30.5$ erg/s; as a 
consequence, with some caution, we can extrapolate the X-ray luminosity function of RASS sources to lower luminosities 
(e.g., $\sim29$ erg/s), and estimate the {\sl total number} of X-ray emitting young stars in the Orion Complex to be 
of the order of a few thousand. 
\item The broad lane apparently connecting the Orion and Taurus SFRs might contain late-type Gould Belt members 
and/or stars belonging to large star halos around Orion and/or Taurus, possibly originated from an earlier episode of 
star formation. We call this component the {\sl dispersed young population} (density of $\sim 0.5-5$ stars/deg$^2$). 
This population appears unrelated to any molecular cloud. The number counts in these areas are in significant excess 
as compared to Galactic model expectations. A high fraction of these sources have strong lithium absorption features rather 
typical of stars with age $\ltsim$ 20 Myr. They also show an iron abundance distribution consistent with that of nearby open 
clusters and associations younger than 150 Myr. Recent analysis including parallax information from {\it Hipparcos} 
and {\it Tycho} suggests that these stars are uniformly distributed along the line of sight, as in the disk-like 
structure proposed by \cite{guilloutetal1998} for the GB, rather than piling up at the supposed outer edge of the Gould 
Belt (\citealt{bally2008}). On the other hand, the interpretation of large star halos around Orion and/or Taurus as result 
of an earlier episode of star formation, say about 20 Myr ago, would require an initial velocity dispersion of nearly 10 
km/s to explain the extension of such halos, whereas the value typically observed for young associations is less than about 
2 km/s, although some authors claim {\sl in situ} star formation in turbulent cloudlets (\citealt{feigelson1996}).
\item The uniformly distributed, {\sl widespread population} has a density $\ltsim$0.5 stars/deg$^2$ near the Galactic Plane 
and is dominated by stars with a lithium abundance compatible with ZAMS or older ages. The number counts agree 
with standard Galactic models. However, a few of these stars appear to be of PMS nature, and explaining their origin remains 
challenging. 
\end{enumerate}
Analogous studies based on RASS data and spectroscopic effort are ongoing in the southern hemisphere and within the 
SACY\footnote{Search for Associations Containing Young stars.} project (\citealt{torresetal2006}). Preliminary results seem 
to show the presence of two populations: $i)$ an old one associated 
with evolved stars (similar to what we call the ``widespread population''); $ii)$ a young one corresponding to 
$\sim$10$-$100~Myr old associations (e.g., $\epsilon$~Cha, Argus, Carina, etc.). 

\section{Conclusions}
\label{sec:conclusions} 
In this paper we analyzed the young stellar content within the ROSAT X-ray All-Sky Survey on 
a scale of several thousand square degrees in the general direction of the Orion star-forming region. 
The study of this young stellar component, through spectroscopic follow-up of a subsample of stars 
and comparison with Galactic model predictions, leads us to the following conclusions: 
{\it The X-ray selected young star candidates consist of a mixture of three distinct populations.}
\begin{enumerate}
\item The youngest {\it clustered population} comprises the dense regions associated with sites of active or 
recent star formation, where the number counts are far in excess with respect to the Galactic model 
predictions. 
\item  The young {\it dispersed population} of late-type stars, whose number counts are in significant excess 
with respect to Galactic model expectations. We cannot firmly establish whether this population represents
the late-type component of the Gould Belt or originated from distinct episodes of star formation.
\item The {\it widespread population} uniformly distributed and dominated by field (ZAMS or older) stars, where 
the number counts are in agreement with standard Galactic models.
\end{enumerate}

In addition, two new young stellar aggregates, namely the ``X-ray clump 0534+22'' and the ``X-ray clump 0430$-$08'', 
were singled out. These aggregates deserve further investigation for their complete characterization. 

Finally, the future {\it Gaia}\footnote{Gaia (Global Astrometric Interferometer for Astrophysics) is an ESA mission that 
will make the largest, most precise three dimensional map by surveying about one billion stars in our Galaxy and 
throughout the Local Group.} mission will provide trigonometric parallaxes and radial velocities for targets in the general 
direction of Orion with a precision of a few parsecs at the Orion distance. Knowing the distances to  
individual stars will permit to place them on the HR diagram with unprecedented precision, allowing us 
to firmly establish the origin of the widespread population of young stars on a Galactic scale.

\begin{acknowledgements} 
This research made use of the SIMBAD database, operated at the CDS (Strasbourg, France). We acknowledge the use of the 
ESO Science Archive Facility. KB acknowledges the financial support from the INAF Postdoctoral fellowship. We thank Fedor 
Getman for software counsel. 
We thank the anonymous referee for his/her careful reading and useful comments and suggestions. 
Last but not least, KB gives special thanks to AM-LP-MZ for their moral support during the preparation of the manuscript,
and JMA \& EC are grateful to MD, RNF, and LMS for encouragement to pursue this work. 
\end{acknowledgements}

\bibliographystyle{aa}

\newpage
\topmargin 7 cm
\pagestyle{empty}

\setlength{\tabcolsep}{3pt}
\begin{landscape}
\tiny
\begin{longtable}{lllc|cc|ccc|lrrl|rcrc|l}
\caption[ ]{\label{tab:parameters} Identified targets with stellar parameters derived through low-resolution and high-resolution spectroscopy.}\\
\hline\hline
~\\
\multicolumn{18}{c}{{\sc ~~~~~~~~~~~~~~~~~~~~~~~~~~~~~~~~~~~~~~~~~~~~~~~~~~~~~~~~~~~~~~~~~~~~~~~~~~~~~~~~~~~~~~~~~~~~~~~~~~~~~~~~~~~~~~~~~~~~~~~~~~~~~~~~~~~~~~~~~~~~~~~~~~~~~~~~~~~~~~~~~~~~~~~~~~~~~~~~~~~~~~~~~~~~~~~~~~~~~~~~~~~LOW~~~~~~RESOLUTION~~~~~~~~~~~~~~~~~~~HIGH~~~~~~~~~~~RESOLUTION}}\\
~\\
\hline
Seq  & 2MASS~J  & 1RXS~J & Other name & $\alpha$ & $\delta$ & $J$ & $H$ & $K$ & Sp.T. & $EW_{\rm Li}^{\rm lr}$ & $EW_{\rm H\alpha}$ & $\log T_{\rm eff}$  & $EW_{\rm Li}^{\rm hr}$ & $\log n{\rm (Li)}$& $V_{\rm rad}$ & $v\sin i$& Note\\
     & 	     &        & 	   & (h:m:s)  & (\degr:$^\prime$:\arcsec) & (mag) & (mag) & (mag) & & (\AA) & (\AA)	   &	(K)		 & (m\AA)	 & (dex) & (km/s) & (km/s)&  \\
\hline
\endfirsthead
\caption{continued.}\\
\hline\hline
~\\
\multicolumn{18}{c}{{\sc ~~~~~~~~~~~~~~~~~~~~~~~~~~~~~~~~~~~~~~~~~~~~~~~~~~~~~~~~~~~~~~~~~~~~~~~~~~~~~~~~~~~~~~~~~~~~~~~~~~~~~~~~~~~~~~~~~~~~~~~~~~~~~~~~~~~~~~~~~~~~~~~~~~~~~~~~~~~~~~~~~~~~~~~~~~~~~~~~~~~~~~~~~~~~~~~~~~~~~~~~~~~LOW~~~~~~RESOLUTION~~~~~~~~~~~~~~~~~~~HIGH~~~~~~~~~~~RESOLUTION}}\\
~\\
\hline
Seq  & 2MASS~J  & 1RXS~J & Other name & $\alpha$ & $\delta$ & $J$ & $H$ & $K$ & Sp.T. & $EW_{\rm Li}^{\rm lr}$ & $EW_{\rm H\alpha}$ & $\log T_{\rm eff}$  & $EW_{\rm Li}^{\rm hr}$ & $\log n{\rm (Li)}$& $V_{\rm rad}$ & $v\sin i$& Note\\
     & 	     &        & 	   & (h:m:s)  & (mag) &(mag)  & (mag) & (\degr:$^\prime$:\arcsec) & & (\AA) & (\AA)	  &   (K)  	      & (m\AA)        & (dex) & (km/s) & (km/s)&  \\
\hline
\endhead
\hline
\endfoot
\hline
~\\
\multicolumn{18}{c}{{\sc ~~~~~~~~~~~~~~~~~~~~~~~~~~~~~~~~~~~~~~~~~~~~~~~~STRIP~~~~~~~~~~~~~~~~~~~~~~~~~~~~~~~~~~~~~~~~~~~~~~~~}}\\
~\\
\hline
1 &03134743$-$1701560 &050927.0+054145	     &       BD-17~625     & 03:13:47.439 &$-$17:01:56.09 &  9.145(0.023) &  8.560(0.057) &  8.449(0.021) &    K0 &    0     &    $-$0.50 &  3.712 &  ...   & ...  &	   ...  &	...  &  	  \\
2 &03214965$-$1052179 &032149.6$-$105228      &       BD-11~648     & 03:21:49.659 &$-$10:52:17.95 &  9.837(0.026) &  9.380(0.026) &  9.264(0.023) &    G9 &    0.250 &       0.40 &  3.719 & 285(10)& 3.07 &   16.1(0.9)&   23.4(1.4)& {\bf a}    \\ 
3 &03494386$-$1353108 &034945.1$-$135315      &       HD~24091	   & 03:49:43.862 &$-$13:53:10.86 &  7.553(0.034) &  7.185(0.031) &  7.070(0.026) &    G9 &    0.150 &       3.30 &  3.719 & 158(5) & 2.52 &   10.9(1.4)&    8.3(1.0)&  	    \\ 
4 &03501928$-$0726234$^\bullet$&035019.3$-$072607&		   & 03:50:19.281 &$-$07:26:23.49 &  9.538(0.026) &  9.116(0.025) &  8.911(0.021) &    K0 &    0.420 &       0.90 &  3.712 & 345(10)& 3.29 &   11.1(1.0)&   28.8(1.4)&  	  \\	 
5 &03503081$-$1355296$^\bullet$&035030.9$-$135527&    BD-14~758     & 03:50:30.818 &$-$13:55:29.63 &  9.061(0.027) &  8.626(0.046) &  8.440(0.020) &    G9 &    0.220 &       0.70 &  3.719 & 145(10)& 2.45 &	...	&   ...      & SB2, {\bf b}, {\bf c}, {\bf d}\\  
6 &03563745$-$1327195$^\bullet$&035637.8$-$132719&		   & 03:56:37.457 &$-$13:27:19.54 &  9.874(0.023) &  9.372(0.022) &  9.194(0.019) &    K2 &    0.450 &    $-$0.50 &  3.685 & 415(10)& 3.25 &	8.5(1.6)&   60.1(0.8)& {\bf c}      \\  
7 &04001941$-$1200583 &040019.2$-$120026      &  		   & 04:00:19.414 &$-$12:00:58.31 & 12.152(0.030) & 11.670(0.029) & 11.571(0.026) &    G5 &    0.180 &       3.50 &  3.745 &  ...   & ...  &	   ...  &      ...   &  	   \\
8 &04081040$-$0749034 &040810.3$-$074854      &       HD~26164	   & 04:08:10.404 &$-$07:49:03.48 &  7.653(0.021) &  7.293(0.033) &  7.214(0.023) &    G2 &    0.150 &       3.05 &  3.763 &  ...   & ...  &	   ...  &      ...   &  	   \\ 
9 &04150615$-$0331515 &041505.4$-$033144      &  		   & 04:15:06.151 &$-$03:31:51.52 & 10.125(0.024) &  9.751(0.023) &  9.679(0.021) &    G9 &    0.270 &       1.32 &  3.719 & 245(10)& 2.89 &   11.3(1.0)&   22.1(1.5)&  	\\
10&04231428$-$0248265 &042314.8$-$024813      &  		   & 04:23:14.287 &$-$02:48:26.54 & 10.531(0.024) &  9.886(0.027) &  9.689(0.023) &    M0 &    0     &    $-$0.60 &  3.584 &  ...   & ...  &	   ...  &      ...   &  	   \\
11&04331165$-$0442029$^\bullet$&043311.6$-$044200&		   & 04:33:11.655 &$-$04:42:02.97 & 10.983(0.023) & 10.396(0.026) & 10.239(0.023) &    K3 &    0.570 &    $-$0.58 &  3.671 & 485(10)& 3.36 &   11.7(0.9)&   12.7(2.3)&  	\\
12&04344277$-$0857184 &043440.8$-$085731      &  		   & 04:34:42.771 &$-$08:57:18.42 & 10.624(0.027) & 10.215(0.026) & 10.126(0.024) &    K1 &    0.200 &       1.60 &  3.698 &  ...   & ...  &	   ...  &      ...   &  	   \\
13&04344238$-$0857223 &043440.8$-$085731      &  		   & 04:34:42.390 &$-$08:57:22.35 & 11.652(0.032) & 10.971(0.029) & 10.840(0.026) &    M0 &    0     &    $-$2.22 &  3.584 &  ...   & ...  &	   ...  &      ...   &  		\\
14&04382712$-$0244070$^\bullet$&043825.9$-$024419&		   & 04:38:27.125 &$-$02:44:07.02 & 10.075(0.024) &  9.621(0.024) &  9.480(0.021) &    K1 &    0.300 &       0.50 &  3.698 & 264(10)& 2.76 &   23.2(1.0)&   24.9(0.7)&  	   \\ 
15&04401624$-$0402238 &044016.2$-$040238      &  		   & 04:40:16.245 &$-$04:02:23.81 & 10.633(0.024) & 10.097(0.023) &  9.900(0.021) &    K6 &    0     &    $-$0.37 &  3.631 &  ...   & ...  &	   ...  &      ...   &  		\\ 
16&04421860+0117399   &044219.2+011741	     &       V1330~Tau     & 04:42:18.609 &$+$01:17:39.94 &  9.666(0.024) &  9.092(0.024) &  8.912(0.019) &    K5 &    0     &    $-$0.25 &  3.644 &  ...   & ...  &	   ...  &      ...   & {\bf e}, RS Var.\\ 
17&04443859$-$0724378 &044437.9$-$072439      &       BD-07~888     & 04:44:38.595 &$-$07:24:37.87 &  8.332(0.024) &  8.033(0.046) &  7.968(0.021) &    G2 &    0.200 &       2.75 &  3.763 &  ...   & ...  &	   ...  &      ...   &  	     \\
18&04500470+0150425   &045005.7+015101	     &       V1831~Ori     & 04:50:04.700 &$+$01:50:42.56 & 10.184(0.022) &  9.734(0.024) &  9.598(0.023) &    G9 &    0.270 &       0.48 &  3.719 & 290(10)& 3.12 &   24.5(1.3)&   66.8(5.7)& SB1, {\bf f} \\ 
19&04554346+0205159$^\bullet$&045543.2+020507 &  		   & 04:55:43.465 &$+$02:05:16.00 & 10.007(0.023) &  9.581(0.022) &  9.475(0.021) &    G6 &    0.310 &       1.35 &  3.738 & 315(10)& 3.41 &   12.6(1.5)&   21.7(2.0)&  	\\
20&05045754$-$0354527 &050457.8$-$035451      &      HD~32704	   & 05:04:57.545 &$-$03:54:52.75 &  6.865(0.024) &  6.378(0.023) &  6.259(0.020) &    G9 &    0.070 &       1.00 &  3.719 &   ...  & ...  &	   ...  &   ...      &       \\
21&05055785+0323326   &050558.2+032331	     &  		   & 05:05:57.857 &$+$03:23:32.67 & 10.903(0.026) & 10.489(0.028) & 10.359(0.025) &    G9 &    0.300 &       1.35 &  3.719 & 235(10)& 2.89 &   16.8(1.5)&   19.3(2.0)&  	  \\
22&05091558+0343483   &050915.7+034339	     &  		   & 05:09:15.589 &$+$03:43:48.33 & 11.112(0.023) & 10.499(0.023) & 10.335(0.019) &    K7 &    0.550 &    $-$1.55 &  3.618 & 475(10)& 2.52 &   17.2(1.5)&   25.0(2.0)&        \\
23&05103912+0554261$^\bullet$&051038.6+055432 &  		   & 05:10:39.129 &$+$05:54:26.15 & 10.254(0.022) &  9.654(0.029) &  9.485(0.021) &M0$^\ast$&  ...   &$-$0.35$^\ast$&3.584$^\ast$&435(5)&1.93& 21.4(1.4)& 22.2(2.0)  &    \\
24&05113738+0255163$^\bullet$&051136.5+025457 &  		   & 05:11:37.381 &$+$02:55:16.36 & 12.675(0.026) & 12.395(0.024) & 12.298(0.027) &    F8 &    0.180 &       5.10 &  3.796 &  ...   & ...  &	   ...  &	...  &  	    \\
25&05133474+0554377$^\bullet$&051334.2+055449 &  		   & 05:13:34.744 &$+$05:54:37.77 &  9.733(0.024) &  9.343(0.024) &  9.224(0.022) &    G8 &    0.360 &       1.15 &  3.725 & 295(5) & 3.19 &   17.8(2.0)&   38.8(2.0)&      \\
26&05274490+0313161   &052744.2+031329	     &  		   & 05:27:44.901 &$+$03:13:16.15 & 13.610(0.027) & 12.955(0.031) & 12.896(0.034) &    K0 &    0.380 &    $-$0.05 &  3.712 & 412(5) & 3.55 &   10.9(1.4)&  14.3(1.5) &      \\
27&05281274$-$0131248 &052744.2+031329	     &  		   & 05:28:12.744 &$-$01:31:24.87 & 10.806(0.023) & 10.555(0.022) & 10.490(0.025) &    F8 &    0.150 &       3.45 &  3.796 &  ...   & ...  &	  ...	&	...  &         \\
28&05281224$-$0131156 &052819.1$-$013011      &  		   & 05:28:12.250 &$-$01:31:15.61 & 11.558(0.024) & 11.157(0.022) & 11.092(0.025) &    G0 &    0.340 &       1.45 &  3.774 &  ...   & ...  &	  ...	&	...  &         \\
29&05303519+0220423   &053035.9+022050	     &  		   & 05:30:35.197 &$+$02:20:42.33 & 10.577(0.022) & 10.177(0.021) & 10.063(0.021) &    K0 &    0.420 &       1.60 &  3.712 & 335(10)& 3.22 &   20.3(1.4)&   20.7(2.0)&        \\
30&05384831+0923557   &053847.7+092407	     &  		   & 05:38:48.319 &$+$09:23:55.80 & 10.482(0.024) & 10.138(0.022) & 10.016(0.019) &    G8 &    0.230 &       1.95 &  3.725 &  ...   & ...  &	   ...  &	...  &  	    \\
31&05445517+1035142   &054454.3+103513	     &  		   & 05:44:55.179 &$+$10:35:14.26 & 10.233(0.022) &  9.843(0.027) &  9.715(0.023) &    K0 &    0.210 &       1.05 &  3.712 &  ...   & ...  &	   ...  &	...  &  	     \\
32&05445274+1035173   &054454.3+103513	     &  		   & 05:44:52.740 &$+$10:35:17.35 & 12.155(0.022) & 11.906(0.024) & 11.754(0.025) &    F9 &    0.160 &       5.40 &  3.785 &  ...   & ...  &	   ...  &	...  &  	       \\ 
33&05540858+1219358   &055408.6+121924	     &       HD~248994     & 05:54:08.581 &$+$12:19:35.89 &  8.954(0.018) &  8.671(0.034) &  8.550(0.016) &    G0 &    0.160 &       2.45 &  3.774 & 146(10)& 3.06 &   38.0(1.4)&   16.2(1.5)&  	    \\ 
34&06030836+0352208   &060308.6+035218	     &  		   & 06:03:08.370 &$+$03:52:20.86 & 11.256(0.024) & 10.855(0.022) & 10.729(0.019) &    K5 &    0.340 &       0.50 &  3.644 & 350(10)& 2.37 &   27.6(4.2)&   58.8(2.0)& SB?  \\ 
35&06030764+0352112   &060308.6+035218	     &  		   & 06:03:07.650 &$+$03:52:11.24 & 11.775(0.024) & 11.277(0.022) & 11.129(0.021) &    G9 &    0.320 &       1.40 &  3.719 &  ...   & ...  &	   ...  &	...  &  	     \\ 
36&06043473+0508489$^\bullet$&060434.3+050902 &  		   & 06:04:34.732 &$+$05:08:48.91 &  9.917(0.022) &  9.536(0.028) &  9.461(0.021) &    G8 &    0.270 &       1.80 &  3.725 & 230(5) & 2.90 &   17.5(1.4)&   21.5(2.0)&  	    \\ 
37&06055511+1228499   &060555.7+122842	     &  		   & 06:05:55.112 &$+$12:28:49.92 & 10.138(0.020) &  9.462(0.023) &  9.259(0.019) &    M1 &    0     &    $-$2.90 &  3.564 &  ...   & ...  &	  ...	&	...  &  	     \\
38&06124499+0944227$^\bullet$&061244.8+094422 &  		   & 06:12:44.995 &$+$09:44:22.71 &  8.579(0.026) &  7.997(0.027) &  7.813(0.029) &    K0 &    0     &       0.28 &  3.712 &  ...   & ...  &	  ...	&	...  &  	     \\ 
39&06134815+0846160$^\bullet$&061347.5+084617 &  		   & 06:13:48.153 &$+$08:46:16.01 &  9.536(0.024) &  9.011(0.023) &  8.891(0.021) &    K2 &    0.420 &       0.75 &  3.685 & 335(10)& 2.86 &   35.7(2.0)&   61.6(2.0)&  	      \\ 
40&06134773+0846022   &061347.5+084617	     &  		   & 06:13:47.738 &$+$08:46:02.25 &  9.555(0.024) &  9.022(0.023) &  8.915(0.023) &    K1 &    0.270 &    $-$0.89 &  3.698 & ...    & ...  &   16.4(1.3)&    5.4(0.5)&  	  \\
41&06154049+1315332   &061540.0+131537	     &       HD~254104     & 06:15:40.490 &$+$13:15:33.30 &  9.630(0.018) &  9.492(0.021) &  9.413(0.017) &    F3 &    0.090 &       5.50 &  3.840 &  ...   & ...  &	   ...  &	...  &  	\\
42&06153978+1315421   &061540.0+131537	     &  		   & 06:15:39.787 &$+$13:15:42.20 & 11.160(0.019) & 10.752(0.021) & 10.669(0.018) &    G9 &    0.240 &       1.65 &  3.719 &  ...   & ...  &	   ...  &	...  &  	\\
43&06203205+1331125$^\bullet$&062031.8+133107 &       HD~255438     & 06:20:32.057 &$+$13:31:12.58 &  8.178(0.027) &  7.921(0.017) &  7.854(0.016) &   ... &    ...   &       ...  &   ...  & 0      & ...  &   21.9(0.9)&    9.6(0.5)&  	   \\
44&06261917+1214546   &062617.3+121510	     &  		   & 06:26:19.178 &$+$12:14:54.61 & 10.170(0.020) & 10.015(0.020) &  9.971(0.017) &    F1 &    0.050 &       6.00 &  3.856 &  ...   & ...  &	   ...  &	...  &  	    \\
45&06261740+1215180   &062617.3+121510	     &  		   & 06:26:17.406 &$+$12:15:18.06 & 10.978(0.021) & 10.404(0.023) & 10.223(0.018) &    K6 &    0.370 &    $-$3.10 &  3.631 & 300(30)& ...  &   18.0(2.3)&   89.5(1.6)&  	  \\	   
46&06300125+1625225   &063000.9+162524	     &  		   & 06:30:01.255 &$+$16:25:22.52 &  9.168(0.022) &  8.715(0.021) &  8.539(0.017) &   ... &    ...   &        ... &  ...   & 0      & ...  &   22.7(1.5)&   55.4(1.5)&  \\
47&06350191+1211359& $^\ddagger$&  		   & 06:35:01.911 &$+$12:11:35.93 & 10.408(0.024) & 10.168(0.025) & 10.101(0.018) &   ... &    ...   &        ... &  ...   &  80(10)& ...  &   37.9(0.8)&    6.7(0.1)&  	       \\ 
48&06382516+1240517& $^\ddagger$&  		   & 06:38:25.169 &$+$12:40:51.76 & 10.297(0.023) &  9.638(0.022) &  9.443(0.018) &    K6 &    0     &    $-$0.90 &  3.631 &  ...   & ...  &   ...	&  ...       &  	      \\ 
49&06413601+0802055$^\bullet$&064136.2+080218 &       HD 262113     & 06:41:36.010 &$+$08:02:05.54 &  9.263(0.026) &  8.781(0.051) &  8.712(0.023) &K1$^\ast$&  0.160 &       1.40&3.698$^\ast$& 160(15)& 2.28 &   19.9(1.4)&	  45.1(2.0)& SB2?, {\bf g}\\ 
50&06453046+1507432   &064529.9+150757	     &  		   & 06:45:30.462 &$+$15:07:43.29 & 10.878(0.022) & 10.419(0.028) & 10.251(0.021) &   ... &    ...   &       ...  &  ...   & 0      & ...  &   58.7(2.7)& $>$100.0   & SB? \\
51&06471556+1434392$^\bullet$&064715.8+143435 &  		   & 06:47:15.568 &$+$14:34:39.27 &  8.850(0.020) &  8.302(0.029) &  8.133(0.026) &    K1 &    0.100 &       0.25 &  3.698 &  ...   & ...  &	   ...  &   ...      &  	   \\ 
52&06513955+1828080$^\Diamond$&065140.3+182801&    TYC~1335-648-1   & 06:51:39.551 &$+$18:28:08.08 &  7.886(0.021) &  7.395(0.027) &  7.291(0.020) &   ... &    ...   &       ...  &  ...   & 0      & ...  & $-$7.4(0.9)&    7.5(0.7)&    \\ 
53&06515913+0936218$^\bullet$&065158.8+093630 &  		   & 06:51:59.137 &$+$09:36:21.87 &  9.491(0.021) &  8.978(0.024) &  8.858(0.023) &    G8 &    0.350 &    $-$1.80 &  3.725 & 350(10)& 3.45 &   24.3(0.9)&   22.3(1.1)&  	    \\
54&06545126+1611078   &065450.7+161112	     &  		   & 06:54:51.267 &$+$16:11:07.87 &  9.854(0.022) &  9.311(0.028) &  9.236(0.020) &   ... &    ...   &       ...  &  ...   &  80(10)& ...  & $-$5.0(0.8)&   32.5(0.5)&  	  \\
55&06550482+2018560$^\bullet$&065505.3+201855 &       HD~266141     & 06:55:04.825 &$+$20:18:56.03 &  8.412(0.021) &  8.192(0.017) &  8.134(0.024) &   ... &    ...   &       ...  &  ...   &  55(10)& ...  &$-$12.0(1.5)&   59.6(1.6)&  \\
56&07002586+1643084   &070026.0+164325	     &  		   & 07:00:25.866 &$+$16:43:08.45 &  9.586(0.029) &  9.107(0.030) &  8.962(0.022) &K3$^\ast$&  ...   &1.20$^\ast$&3.671$^\ast$&165(10)&1.98&   91.7(0.9)&   26.3(1.4)&  \\
57&07011843+1528366   &070118.7+152849	     &  		   & 07:01:18.439 &$+$15:28:36.61 & 10.729(0.024) & 10.138(0.030) & 10.038(0.019) &    K1 &    0     &    $-$0.55 &  3.698 &  ...   & ...  &	   ...  &      ...   &  	     \\
58&07023741+1558261$^\bullet$&070237.1+155843 &        HD~52634     & 07:02:37.414 &$+$15:58:26.17 &  7.158(0.023) &  6.918(0.038) &  6.873(0.020) &    F9 &    0.170 &       3.28 &  3.785 & 150(10)& 3.17 &   25.8(1.4)&   22.2(2.0)& {\bf h}, Doub. Syst.	\\ 
59&07071097+2011394   &070710.2+201136	     &  TYC~1353-331-1     & 07:07:10.974 &$+$20:11:39.41 &  9.289(0.024) &  9.003(0.024) &  8.958(0.018) &    F9 &    0.060 &       3.90 &  3.785 &  ...   & ...  &	   ...  &      ...   &        \\ 
60&07110918+1312442   &071109.2+131246        &			   & 07:11:09.183 &$+$13:12:44.24 &  8.644(0.027) &  8.014(0.018) &  7.848(0.023) &    M2 &    0     &       0.18 &  3.547 &  ...   & ...  &	 ...  &	     ...   & 	 \\ 
61&07181093+1735160$^\bullet$&071811.4+173515 &			   & 07:18:10.930 &$+$17:35:16.01 &  8.592(0.026) &  8.111(0.018) &  7.978(0.027) &    K0 &    0.300 &       1.20 &  3.712 & 350(30)& 3.29 &       ...  &	     ...   &	\\ 
\hline	
~\\	
\multicolumn{18}{c}{{\sc ~~~~~~~~~~~~~~~~~~~~~~~~~~~~~~~~~~~~~~~~~~~~~~~~X-CLUMP 0534+22~~~~~~~~~~~~~~~~~~~~~~~~~~~~~~~~~~~~~~~~~~~~~~~~}}\\
~\\
\hline	
62&05203710+2447135$^\bullet$&052036.6+244731  &      V1360~Tau   & 05:20:37.104 &$+$24:47:13.54&  9.767(0.020) &  9.257(0.022) &  9.072(0.017) &    K5 &      0.370 &$\sim 0.0$ &  3.644 & 435(10)& 2.73 &   20.7(1.2)& 103.3(11.3)& RS Var.   \\ 
63&05214684+2400444$^\bullet$&052146.7+240036  &      V1361~Tau   & 05:21:46.844 &$+$24:00:44.43&  8.592(0.026) &  8.111(0.018) &  7.978(0.027) &    K3 &      0.350 &	$-$0.40 &  3.671 & 395(5) & 2.93 &   14.2(1.4)& 15.6(1.0)  & {\bf i}, T Tau   \\ 
64&05221036+2432089$^\bullet$&052210.2+243200  &      V1362~Tau   & 05:22:10.360 &$+$24:32:08.96&  8.556(0.023) &  7.991(0.018) & 7.889$^\star$ &    G9 &      0.330 &	$-$0.25 &  3.719 & 365(10)& 3.45 &   19.1(1.4)& 26.8(2.0)  & RS Var.	       \\
65&05224717+2437311$^\bullet$&052248.0+243731  &      V1363~Tau   & 05:22:47.171 &$+$24:37:31.13&  9.268(0.024) &  8.670(0.031) &  8.480(0.023) &    K4 &      0.490 &	$-$0.20 &  3.657 & 475(10)& 3.08 &   21.4(1.9)& 46.4(2.0)  & RS Var. \\
66&05263833+2231546$^\bullet$&052638.7+223151  & 		 & 05:26:38.334 &$+$22:31:54.66& 10.119(0.022) &  9.536(0.022) &  9.415(0.020) &    K5 &      0.280 &	$-$0.04 &  3.644 & 290(5) & 2.15 &   28.0(1.4)& 14.4(1.0)  & T Tau\\
67&05263826+2231434$^\bullet$&052638.7+223151  & 		 & 05:26:38.269 &$+$22:31:43.48& 10.001(0.021) &  9.478(0.022) &  9.340(0.017) &    K6 &      0.290 &	$-$0.14 &  3.631 & 285(10)& 1.91 &   28.3(1.4)& 10.0(1.0)  & T Tau     \\
68&05270306+2041508 &052703.5+204204	      & 		 & 05:27:03.066 &$+$20:41:50.86&  9.255(0.021) &  8.542(0.026) &  8.410(0.017) & K7-M0 &      0.400 &	$-$1.20 &  3.595 & 447(10)& 2.12 &   12.8(1.4)& 12.7(1.0)  &		    \\
69&05271996+2503434 &052720.0+250348	      & 		 & 05:27:19.962 &$+$25:03:43.46& 10.445(0.023) &  9.757(0.022) &  9.547(0.018) &    M0 &      0.430 &	$-$1.50 &  3.584 & 515(10)& 2.24 &   17.5(1.4)& 11.3(1.0)  &		    \\
70&05322227+2521077$^\Diamond$&053222.9+252106 &  TYC~1852-1665-1 & 05:32:22.279 &$+$25:21:07.77&  9.029(0.021) &  8.683(0.022) &  8.603(0.019) &    G7 &      0     &	   2.23 &  3.732 &  ...   & ...  &	 ...  &       ...  &		   \\ 
71&05332381+2019575$^\bullet$&053323.5+201951  &   TYC~1305-353-1 & 05:33:23.816 &$+$20:19:57.50&  8.789(0.023) &  8.328(0.024) &  8.270(0.020) &    K0 &      0     &	   1.55 &  3.712 &  ...   & ...  &	 ...  &       ...  &	  \\
72&05394828+2614008 &053948.9+261427	      & 		 & 05:39:48.287 &$+$26:14:00.84& 10.054(0.020) &  9.678(0.022) &  9.561(0.017) &   ... &      ...   &     ...  &   ...  & 30(10)&  ... & 34.5(2.2)& 19.3(1.0)  &     \\ 
73&05410142+2036179$^\bullet$&054101.8+203624  & 		 & 05:41:01.429 &$+$20:36:17.96&  9.160(0.021) &  8.659(0.021) &  8.504(0.018) &    K6 &      0.300 &	$-$1.20 &  3.631 & 375(10)& 2.31 &   34.5(1.2)& 76.9(4.7)  &	     \\ 
74&05463252+2435382 &054632.7+243549	      & 		 & 05:46:32.524 &$+$24:35:38.25&  9.651(0.021) &  9.200(0.021) &  9.068(0.017) &    K4 &      0.300 &	   0.90 &  3.657 & 380(5) & 2.67 &   19.9(1.4)& 24.2(2.0)  &		      \\ 
75&05463283+2240315 &054632.8+224041	      & 		 & 05:46:32.831 &$+$22:40:31.54&  9.685(0.021) &  9.031(0.024) &  8.896(0.020) &    K6 &      0.400 &	$-$0.20 &  3.631 & 315(10)& 2.03 &   15.4(1.4)& 12.8(1.0)  &		      \\
76&06020094+1955290 & $^\ddagger$& 		 & 06:02:00.942 &$+$19:55:29.02& 11.269(0.023) & 10.705(0.032) & 10.546(0.020) & K7-M0 &      0.170 &	$-$0.75 &  3.595 & 130(10)& 0.77 &   13.4(1.3)& 22.9(0.1)  &        \\ 
\hline  
~\\	
\multicolumn{18}{c}{{\sc ~~~~~~~~~~~~~~~~~~~~~~~~~~~~~~~~~~~~~~~~~~~~~~~~L1616 CLUMP~~~~~~~~~~~~~~~~~~~~~~~~~~~~~~~~~~~~~~~~~~~~~~~~}}\\
~\\
\hline	
77&04591458$-$0337062$^\Diamond$&045912.4$-$033711&	  & 04:59:14.023 &  $-$03:37:06.08    & 10.070(0.022) &  9.616(0.024) & 9.474(0.020) &      G7 &     0.340 &   0.35 &	   3.732 & 355(10)& 3.53 &   12.2(1.6)&   32.4(2.0)& {\bf g}, {\bf m}, {\bf n}, T Tau\\
78&05041593$-$0214505$^\Diamond$&050416.9$-$021426&	  & 05:04:15.932 &  $-$02:14:50.51    & 10.661(0.024) & 10.105(0.024) & 9.984(0.025) &      K3 &     0.490 &$-$0.18 &	   3.671 & 475(10)& 3.30 &   19.8(1.5)&   29.1(2.0)& {\bf m}, {\bf n}, T Tau\\
79&05090066$-$0315066$^\bullet$&050859.6$-$031503&V1849~Ori& 05:09:00.662 &  $-$03:15:06.63    &  9.914(0.024) &  9.530(0.024) & 9.408(0.021) &K1$^\ast$&...  &   1.29$^\ast$&3.698$^\ast$& 320(10)& 2.99 &   23.4(2.0)&   40.6(2.0)& {\bf c}, {\bf l}, {\bf m}, {\bf n}, T Tau \\
80&05101086$-$0254049$^\Diamond$&051011.5$-$025355&V1011~Ori&05:10:10.860 &  $-$02:54:04.94    & 10.454(0.022) &  9.952(0.024) & 9.730(0.023) &      K1 &     0.480 &   0.35 &	   3.698 & 415(10)& 3.45 &   26.1(1.5)&   31.2(2.0)& {\bf m}, {\bf n}, T Tau	      \\ 
81&05101478$-$0330074 &051015.7$-$033001 &		  & 05:10:14.783 &  $-$03:30:07.40    & 10.038$^\star$&  9.806(0.060) & 9.749(0.049) &G9$^\ast$&...  &   1.10$^\ast$&3.719$^\ast$& 320(10)& 3.24 &   19.0(1.8)&   34.6(2.0)& {\bf l}, {\bf m}, {\bf n}, T Tau \\
82&05104050$-$0316415 &051043.2$-$031627 &  TYC~4755-873-1 & 05:10:40.504 &  $-$03:16:41.56    & 10.079(0.026) &  9.735(0.022) & 9.648(0.025) &      G5 &     0.240 &   2.70 &	   3.745 & 235(10)& 3.20 &   20.1(2.6)&   78.9(2.0)& {\bf m}, {\bf n}, T Tau		 \\ 
83&05122053$-$0255523 &051219.9$-$025547 &    V531~Ori	  & 05:12:20.531 &  $-$02:55:52.34    & 10.425(0.023) &  9.688(0.023) & 9.140(0.019) &K3$^\ast$&...  &$-$6.35$^\ast$&3.671$^\ast$& 395(10)& 2.93 &   23.3(1.2)&   34.0(1.8)& {\bf l}, {\bf m}, {\bf n}, Var. Rapid \\ 
\hline	
~\\	
\multicolumn{17}{c}{{\sc ~~~~~~~~~~~~~~~~~~~~~~~~~~~~~~~~~~~~~~~~~~~~~~~~X-CLUMP 0430$-$08~~~~~~~~~~~~~~~~~~~~~~~~~~~~~~~~~~~~~~~~~~~~~~~~}}\\
~\\
\hline	
84&04405981$-$0840023 &044059.2$-$084005 	  &	MM~Eri      & 04:40:59.812 &  $-$08:40:02.38	  &  8.880(0.023) &  8.528(0.026)&8.558(0.027)&G7$^\ast$&...&2.50$^\ast$&3.732$^\ast$& 168(10)& 2.69 &   20.8(1.5)&  29.9(2.0)&  {\bf b}, {\bf c}, RS Var.   \\ 
85&04354055$-$1017293$^\bullet$&043541.2$-$101731  & TYC~5317-3258-1 & 04:35:40.553 &  $-$10:17:29.36	  &  9.620(0.024) &  9.269(0.024)&9.174(0.023)&K1$^\ast$&...&3.50$^\ast$&3.698$^\ast$& 195(10)& 2.47 &   14.5(1.8)&  34.0(2.0)& 		  \\ 
86&04431640$-$0937052 &2E 0440.9$-$0942$^\triangle$&    HD~29980     & 04:43:16.408 &  $-$09:37:05.28	  &  6.971(0.021) &  6.693(0.055)&6.621(0.023)&G5$^\ast$&...&3.40$^\ast$&3.745$^\ast$&  85(5) & 2.45 &   33.5(0.9)&   5.7(0.1)& {\bf h}, Doub. Syst. \\ 
87&04463244$-$0857241 &044632.7$-$085723 	  & TYC~5322-1381-1 & 04:46:32.450 &  $-$08:57:24.20	  &  9.142(0.032) &  8.466(0.026)&8.393(0.024)&K2$^\ast$&...&1.40$^\ast$&3.685$^\ast$&  0     & ...  &$-$10.9(1.3)&  22.0(2.0)&   \\
88&04503013$-$0837103 &045029.9$-$083701 	  &		    & 04:50:30.137 &  $-$08:37:10.39	  & 10.581(0.024) & 10.048(0.024)&9.900(0.021)&K4$^\ast$&...&0.27$^\ast$&3.732$^\ast$& 210(10)& 2.92 &   16.0(0.9)&  15.0(0.4)& 	  \\
89&04443859$-$0724378 &044437.9$-$072439 	  &    BD-07~888    & 04:44:38.595 &  $-$07:24:37.87	  &  8.332(0.024) &  8.033(0.046)&7.968(0.021)&G7$^\ast$&...&3.30$^\ast$&3.732$^\ast$& 160(5) & 2.66 &   23.1(0.9)&   3.4(0.1)& 	   \\
90&04383054$-$0645583$^\bullet$&043830.8$-$064559  & TYC~4747-376-1  & 04:38:30.548 &  $-$06:45:58.36	  &  9.608(0.029) &  9.274(0.026) & 9.132(0.026) & G0 & 0.210	 & 2.55 &      3.774 & 210(10)& 3.37 &    3.8(1.5)&  22.5(2.0)& \\
91&04390790$-$0805581 &043905.9$-$080619 	  &		    & 04:39:07.902 &  $-$08:05:58.11	  & 10.202(0.027) &  9.747(0.022) & 9.614(0.021) & G9 & 0.410	 & 1.10 &      3.719 & 350(10)& 3.38 &   14.7(1.7)&  36.4(2.0)&   \\
\hline
\end{longtable}
\footnotesize{\noindent{Main references: {\bf a}: \cite{dasilvaetal2009}; {\bf b}: \cite{covinoetal2001}; {\bf c}: \cite{marillietal2007}; {\bf d}: \cite{favataetal1995}; 
{\bf e}: \cite{torresetal2002}; {\bf f}: \cite{broegetal2006}; {\bf g}: \cite{szczygieletal2008}; {\bf h}: \cite{gontcharov2006}; 
{\bf l}: \cite{alcalaetal2000}; {\bf m}: \cite{alcalaetal2004}; {\bf n}: \cite{gandolfietal2008}.}} \\
\footnotesize{Notes:
\begin{itemize}
\item $^\ast$ Values obtained from high-resolution spectra. 
\item $^\ddagger$ Star not present in the ROSAT All-Sky Bright Source Catalogue. 
\item $^\triangle$ Designation of the {\it Einstein} Soft X-ray Source List (\citealt{mcdowell1994}).
\item $^\star$ 2MASS magnitude of low quality.
\item $^\bullet$ Already identified as 2MASS point source by \cite{haakonsenrutledge2009}.
\item $^\Diamond$ Already identified as 2MASS point source by {\it Simbad}.
\item SB1: single-lined spectroscopic binary; SB2: double-lined spectroscopic binary; SB2?: suspected spectroscopic binary; SB?: suspected spectroscopic multiple.
\item {\it Simbad} notes: Susp. Var.: Star suspected of Variability; RS Var.: Variable of RS CVn type; Doub. Syst.: Star in double system; T Tau: T Tau-type Star; Var. Rapid: Variable Star with rapid variations.
\end{itemize}
}
\end{landscape}

\newpage
\topmargin -.2 cm
\pagestyle{empty}

\Online

\begin{appendix}
{\bf To be published in electronic form only}

\section{Large-scale spatial distribution of the targets}

\begin{figure*}[h]
\begin{center}
  \includegraphics[width=17cm]{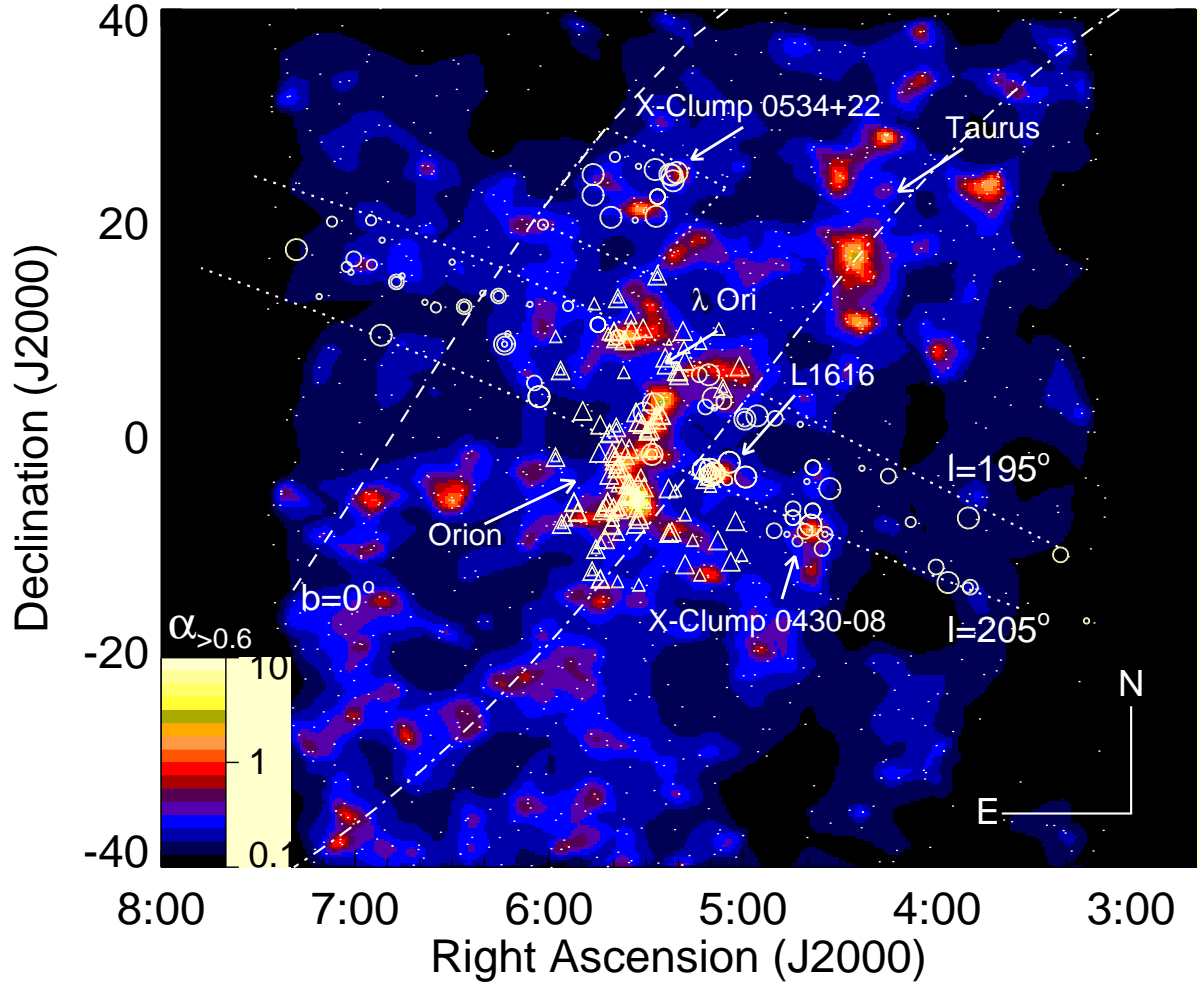}            
\caption{Large-scale spatial distribution of 1483 candidate young stars (small dots) in the Orion and Taurus-Auriga 
associations. The X-ray sources have been selected from the ROSAT All-Sky Survey. Their space density is color-coded, 
where the parameter $\alpha$ represents the discriminant probability; the higher $\alpha$ the more probable is the 
occurrence of a source corresponding to a young star. Circles refer to sources observed spectroscopically in this work, 
while triangles represent targets previously analyzed by \cite{alcalaetal1996} and \cite{alcalaetal2000}. The symbol 
size increases with the lithium equivalent width $EW$ (0, $<$150, 150--300, $>$300 m\AA). The dashed line coincides with 
the position of the Galactic Plane, while the dash-dotted line represents the Gould Belt, as given by \cite{guilloutetal1998}. 
The sky strip ($195{\degr}<l_{II}<205{\degr}$, $-60{\degr}<b_{II}<15{\degr}$) selected by us across the Gould Belt in the 
Orion vicinity is also shown together with the area selected on the X-Clump 0534+22 (see Sect.~\ref{sec:follow-up}). The 
regions of enhanced density of young star candidates and of high-lithium stars southward of the Galactic Plane coincide 
with the location of the Gould Belt midplane. North is up and East to the left.} 
\label{fig:map}
\end{center}
\end{figure*}

\begin{figure*}	
\begin{center}
 \begin{tabular}{c}
\includegraphics[width=15cm]{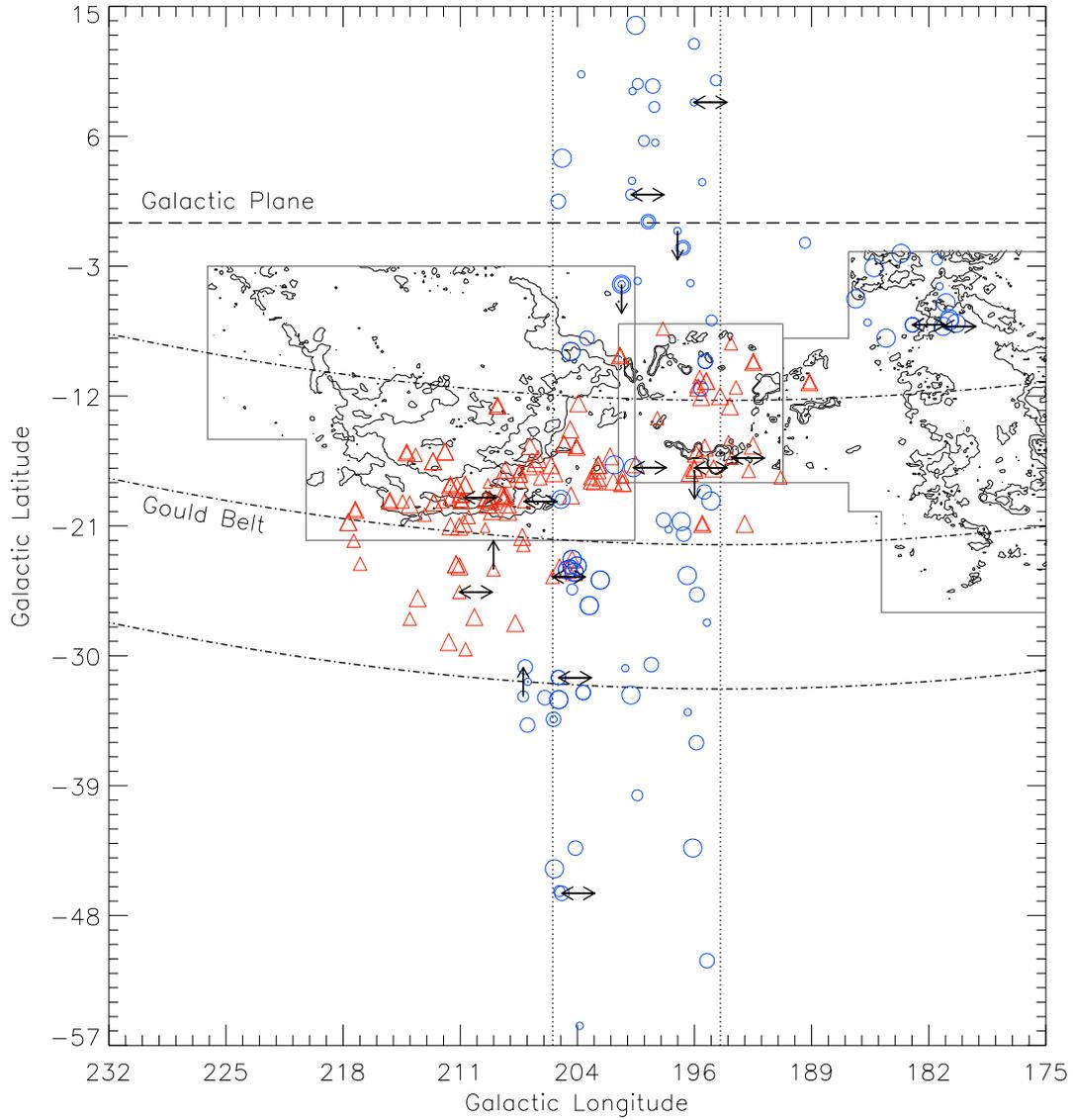}
 \end{tabular}
\vspace{-1cm}
\caption{Large-scale spatial distribution of all targets observed at low and/or high resolution. Circles refer to sources 
observed spectroscopically in this work, while triangles are targets analyzed by \cite{alcalaetal1996} and \cite{alcalaetal2000}. 
The symbol size increases with the lithium equivalent width (0, $<$150, 150--300, $>$300 m\AA). The arrows refer to stars for 
which we measured the iron abundance (downward for [Fe/H]$\le-0.14$, horizontal for $-0.14<$[Fe/H]$\le 0.06$, upward for 
[Fe/H]$>0.06$; the central bin corresponds to the average metallicity of our targets  $\pm 1\sigma$). Dotted lines represent 
the strip, long-dashed line is the Galactic Plane, while dash-dotted lines represent the Gould Belt disk, as outlined by 
\cite{guilloutetal1998}. The CO J=1$\rightarrow$0 emission contour maps by \cite{dameetal2001} of the Orion, Monoceros, 
and Taurus Molecular Clouds, and the $\lambda$~Orionis \ion{H}{ii} region are also overlaid.
}
\label{fig:map_fe}
 \end{center}
\end{figure*}

\end{appendix}

\end{document}